\documentclass[journal]{IEEEtran}

\usepackage{amssymb}
\usepackage{amsmath}
\usepackage{cite}
\usepackage{verbatim}
\usepackage{textcomp}
\usepackage[pdftex]{graphicx}
\usepackage{grffile}
\usepackage{caption}
\usepackage{subcaption}
\usepackage{epsfig}
\usepackage{epstopdf}
\usepackage{caption}
\usepackage[ruled,vlined,linesnumbered, commentsnumbered]{algorithm2e}
\usepackage{balance}
\usepackage{listings}

\SetCommentSty{mycommfont}

\usepackage{tipa }	
\usepackage{pifont}	
\usepackage{wasysym}
\usepackage{color}

\usepackage{setspace}

\usepackage{amsthm}

\newtheorem{definition}{Definition}

\DeclareMathOperator*{\argmax}{arg\,max}
\DeclareMathOperator*{\argmin}{arg\,min}

\usepackage{color}

\begin{document}	

\title{Balanced Wireless Crowd Charging with Mobility Prediction and Social Awareness}

\author{
\IEEEauthorblockN{Tamoghna Ojha, Theofanis P. Raptis, Marco Conti, and Andrea Passarella}\\
\IEEEauthorblockA{Institute for Informatics and Telematics, National Research Council, Italy\\
Email: \small \{tamoghna.ojha, theofanis.raptis, marco.conti, andrea.passarella\}@iit.cnr.it
}}


\maketitle

\begin{abstract}\label{Abst}
The advancements in peer-to-peer wireless power transfer (P2P-WPT) have empowered the portable and mobile devices to wirelessly replenish their battery by directly interacting with other nearby devices. The existing works unrealistically assume the users to exchange energy with any of the users and at every such opportunity. However, due to the users' mobility, the inter-node meetings in such opportunistic mobile networks vary, and P2P energy exchange in such scenarios remains uncertain. Additionally, the social interests and interactions of the users influence their mobility as well as the energy exchange between them. The existing P2P-WPT methods did not consider the joint problem for energy exchange due to user's inevitable mobility, and the influence of sociality on the latter. As a result of computing with imprecise information, the energy balance achieved by these works at a slower rate as well as impaired by energy loss for the crowd. Motivated by this problem scenario, in this work, we present a wireless crowd charging method, namely \textit{MoSaBa}, which leverages mobility prediction and social information for improved energy balancing. MoSaBa incorporates two dimensions of social information, namely social context and social relationships, as additional features for predicting contact opportunities. In this method, we explore the different pairs of peers such that the energy balancing is achieved at a faster rate as well as the energy balance quality improves in terms of maintaining low energy loss for the crowd. We justify the peer selection method in \textit{MoSaBa} by detailed performance evaluation. Compared to the existing state-of-the-art, the proposed method achieves better performance trade-offs between energy-efficiency, energy balance quality and convergence time.

\end{abstract}

\begin{keywords}
Wireless power transfer, peer-to-peer, mobile opportunistic networks, energy balance, mobility
\end{keywords}


\section{Introduction}\label{Intro}
Energy of any mobile device is the most important resource for ensuring its continued functionality. With the increased introduction of applications as well as growing engagement and dependence of users towards their mobile phones, the already `limited' energy resource of the mobile and portable devices is becoming more precious and vital. In this regard, the recent advancements in wireless power transfer (WPT) technologies have empowered the portable and mobile devices to be independent of the wired chargers, and instead use the wireless charging techniques for replenishing their batteries for seamless execution of running applications \cite{Dhungana2020} \cite{Zhang2018}. For example, devices which enable wireless charging process for these mobile devices are wireless charging pads and mobile charging vehicles \cite{Gao2020}, \cite{Li2019}. In the near future, we can expect that the specific advancements in peer-to-peer (P2P) WPT methods will allow any mobile device to replenish its battery by directly interacting and exchanging energy wirelessly from the nearby mobile devices. In this regard, the use of P2P-WPT-based mobile charging techniques provide increased flexibility for the mobile and portable devices to continue exchanging energy while maintaining their mobility. Specifically, in constrained environments, where external chargers might not be available, the P2P-WPT techniques can provide the only practical and effective solution. 

With the rapid developments in WPT standards (e.g. Qi \cite{Qi_standard}), the leading smart-phone manufacturers have incorporated wireless charging standards in their mobile portable devices \cite{qi_phones}. The wireless charging market is also experiencing growth in a rapid pace of 23.15\% with an estimated total market of 7.43 Billion \cite{wpt_market}. Consequently, the adoption of P2P-WPT techniques for various new applications also emerged. Two such applications are wireless powered communication networks and crowd charging. In the first type, the deployed nodes focus on harvesting energy from a hybrid access point and use it to transmit their information \cite{iqbal2019minimum}, \cite{kazmi2020total}. Whereas, in crowd charging, any node of the network can engage in energy exchange with any other node without the presence of any access point. Therefore, the energy exchange in crowd charging is distributed and multiple pairs can engage simultaneously. Typically, in P2P-WPT techniques deployed for such crowd charging applications, an important aspect is to achieve the state of `energy balancing' over the whole network such that all the nodes reach an equal energy level. Energy balancing of any network helps in replenishing the battery levels of nodes that have drained their energy, and thereby, helps in enhancing the functional lifetime of the network. However, such networks also face the problem of uncertain and opportunistic meetings between the nodes due to their mobility. Due to such uncertainty, the inter-meeting duration between the nodes varies, and the energy exchange between the nodes in such cases remain impaired by the varying energy loss (energy losses are inevitable in each WPT interaction, due to the wireless attenuation). Subsequently, the resulting network may have the crowd with unbalanced energy levels and high value of energy variation distance across devices. Therefore, in such cases, to ensure the quality of energy balancing, it is equally important to consider both the amount of energy to be exchanged and the available duration of exchange.

In the existing literature, the P2P-WPT techniques assume nodes to exchange energy at every available opportunity \cite{Nikoletseas2017} with other nodes. Furthermore, few techniques also unrealistically consider all the nodes to be available for energy exchange with all other nodes \cite{Dhungana2019}. The energy exchange duration in such techniques are also not bounded by the inter-meeting duration between the corresponding users \cite{Dhungana2021}. In reality, any node can meet only a fraction of the crowd over the course of movement. As a result, the nodes can not have knowledge about all the other nodes, rather only partial information about the rest of the nodes. Therefore, the selection of peers using distributed knowledge may result in high and varying energy loss as well as low average network energy. In this regard, a centralized peer selection method with knowledge on users' mobility will have advantages due to the availability of information on different pairs of peers and their potential energy loss.

In addition to above challenges, the energy balancing techniques in crowd charging also depend on the social interactions and the interests of the users, which influences the mobility of the users \cite{Karamshuk2011} as well as the energy sharing between them. The existing P2P-WPT techniques only consider social interactions between groups of users or between individual users \cite{Raptis2020} \cite{Bulut2020social}. However, they did not consider the joint problem for energy exchange due to user's inevitable mobility and social influences. Therefore, the energy balancing process in these techniques did not exploit all the opportunities between different pairs of nodes. With a centralized knowledge-based peer selection method having the information about the social interests and interactions of the users, the crowd can intelligently explore the pairs of peers to achieve faster energy balance with lower energy loss and higher energy in the network.

\subsection{Contributions}
In this paper, we present a \underline{mo}bility and \underline{s}ocial-\underline{a}ware energy \underline{ba}lancing method, named $MoSaBa$, which applies centralized knowledge-based peer selection algorithm in P2P-WPT. The objective of the proposed method is to maximize the number of nodes which reaches energy balancing while reducing the energy loss and variation distance. To achieve these objectives, $MoSaBa$ first collects the mobility information of the nodes as well as the social information. We explore two distinct dimensions of social information -- social context and contacts (or, relations) information. Thereafter, an $O(k)$ Markov predictor is applied to predict the future mobility information. Next, in two incremental steps, the peers are selected based on the information on social context and contacts such that the energy loss and variation distance among the crowd is minimized. For social contexts, we incorporate the location-based interests in this model. Whereas, for social contacts, the self-reported social contacts are identified. In summary, the \textit{contributions} made in this paper are as follows,
\begin{itemize}
\item We explore best matching of pairs of peers considering their mobility and social information such that the total network energy after energy balancing is maximized.
\item We leverage the maximum contact among each pair of peers such that the energy loss during the exchange is minimized.
\item The energy balancing quality for the crowd is maximized considering their mobility and social information.
\item In MoSaBa, we incorporate social information as an additional feature for predicting contact opportunities. The two dimensions of social information, namely social context and social relationships, are included incrementally in our work.
\end{itemize}

In our previous work \cite{Ojha2021}, we have addressed the problem of mobility-aware energy balancing for P2P-WPT. In this work, we extend our previous work with scope of exploiting user's interests and social contacts for enhanced peer selection in P2P-WPT. Specifically, our extended work studies the joint problem for energy exchange due to user movement with their social interests as well as interactions between the users and their social friends. Consequently, using the centralized knowledge-based algorithm, we explore further opportunities in peer selection for P2P-WPT with the specified objectives to enhance the energy balance quality.

The rest of the paper is organized as follows. Section \ref{sec:RelWrk} summarizes the related literature in the domain of P2P-WPT. Next, in Section \ref{sec:SysModel}, we describe the network model and key concepts of the proposed method with an example scenario highlighting the main concepts of the method. The proposed mobility and social-aware wireless energy balancing method is presented in Section \ref{sec:Mobi_Soc_WEB}. We explain the performance evaluation scenario and highlight the incremental benefits of the social-related components of the proposed method in Section \ref{sec:Result_1}. Subsequently, in Section \ref{sec:Result_2}, we discuss the state-of-the-art benchmarks and results of the proposed method in comparison to them. Finally, we conclude the paper in Section \ref{sec:Conclu}, pointing directions for future research.

\section{Related Works}\label{sec:RelWrk}
The state-of-the-art on P2P-WPT can be classified into three different categories based on their objectives and working methodology. In the following, we discuss works from each category and highlight the value proposition of our proposed work compared to these.

In the first category, we discuss the works which focus on balancing the available crowd energy while maintaining low energy loss. \cite{Nikoletseas2016a, Nikoletseas2017} considered scenarios with both loss-less and lossy WPT, and computed the upper bound of the time required for the energy balancing of the whole crowd in both the scenarios. Although, the loss-less P2P-WPT is not realistically possible, this study provides a theoretical foundation for analyzing P2P-WPT performance of a distributed crowd. The authors device three different P2P-WPT methods for loss-less and lossy scenarios targeting different objectives such as minimizing energy loss, minimizing time to reach energy balance, and knowledge about other nodes' energy status. In another work \cite{Michizu2018}, Michizu \textit{et al.} proposed a method for balancing the energy of the crowd while minimizing the energy loss in the WPT process. In contrast to the works in this category, in our work, we present a fine-grained realistic P2P-WPT among the users, as well as, we leverage the mobility information of the users to improve the peer selection during WPT.

In the second category, we discuss the works which focus on reshaping the network energy graph of the crowd -- from arbitrary graphs to a well-defined formation. For example, Madhja \textit{et al.} proposed two different method to distributively form a star type network graph \cite{Madhja2016}, and a tree type network graph \cite{Madhja2018}. In similar direction, Bulut \textit{et al.} \cite{Bulut2014}, devised a method for dynamically assigning roles to nodes and changing the graph hierarchy in the process -- some of the nodes were selected to provide energy for the other nodes of the crowd such that the all the nodes of the crows remained functional. Another related work by Bulut \textit{et al.} \cite{Bulut2017} analyzed the benefits of energy sharing between nodes to reduce inefficient utilization of energy. The authors explored the WPT interaction opportunities between the users, and based on it, assigned users to one-another for P2P energy exchange. \cite{Sakai2021} proposed another method which first identifies the important nodes in the network that are well connected with rest of the nodes, and then, these nodes are chosen as power source node for rest of the nodes. However, the assumed power transmission efficiency is high, and therefore, may be less effective is real-life. The proposed work, in contrast to the works of this category, explores both the mobility information of the users and social relations between them for improving the WPT peer selection process.

Finally, for the third category, we discuss the works which focus on joint objectives of energy as well as ICT services. For example, Dhungana \textit{et al.} \cite{Dhungana2019a} proposed a method to enhance content delivery in the network by providing incentives to nodes which carry the content towards delivering at the destination. In another work \cite{Dhungana2019b}, the authors devised a method to reduce the dependency of users on wired charging. The authors proposed to assign users from the crowd for engaging in WPT such that it maximizes the charging relief throughout the crowd. The aspect of social network was also explored for devising strategies for peer selection for P2P-WPT among a crowd. Raptis \cite{Raptis2020} first proposed this idea of socially-motivated energy exchange among the peers, and accordingly, presented two different methods which enable energy exchange between individuals explicitly considering their social relations. Advancing the direction of social influence on energy balancing, Bulut and Dhungana \cite{Bulut2020social} studied the problem in the context of opportunistic networks. The method devised by the authors was targeted for achieving better energy balance in terms of variation distance. However, the social structure considered by the authors remains non-generic to realistic social networks, and the method also incurred additional energy loss. In our proposed work, we consider realistic social structure where individual users can be friends with any other individual. We also consider the realistic meeting between the users influencing the P2P-WPT process between themselves.

\textbf{Inference}: The existing works for energy balance in P2P-WPT exhibit unrealistic, coarse-grained and inefficient peer selection for P2P-WPT leading to higher energy loss and variable energy distribution in the resulting crowd. Additionally, the energy balancing processes in the existing works only partially consider the social dimension influencing the P2P meeting. Subsequently, these methods remain limited in exploring all the P2P meeting opportunities. In comparison to this, the proposed method explores the joint problem for energy exchange due to user's inevitable mobility and social influences.


\section{Network Model and Key MoSaBa Concepts}\label{sec:SysModel}
We assume $m$ users each carrying a mobile device, $\mathcal{U} = \{ u_1, u_2, \cdots, u_m \}$, present over the area of interest $\mathbb{A}$ consisting of $n$ locations $\mathbb{L} = \{ \mathcal{L}_1, \mathcal{L}_2, \cdots, \mathcal{L}_n \}$. The location of any node $u_i$ at time $t$ is denoted as $l^i_t$, and the locations of all the nodes are referred as, $\mathtt{L}(t) = \{l^1_t, l^2_t, \cdots , l^m_t\}$. $\mathcal{E}_t = \{ E_t(1), E_t(2), \cdots, E_t(m) \}$ are the energy levels of the nodes at time $t$. For simplicity, we consider that all the nodes have homogeneous WPT hardware, and thus, have an equal total battery capacity. In this work, we employ one-dimensional abstraction suggested by Friis formula for the power received by one antenna under idealized conditions given another antenna some distance away. However, as mentioned in our previous work \cite{Kat17}, we can apply a more detailed vectorial model which arises naturally from fundamental properties of the superposition of energy fields. However, we note that in large-scale wirelessly powered networks modelling there is always a trade-off to strike between fine-grained modelling accuracy and feasibility of (i) combinatorial problem definition, (ii) algorithmic design and (iii) simulation execution. Our current modelling approach has been widely accepted in the related literature (for example, see \cite{Niko17}, \cite{Dhungana2019}, \cite{Dhungana2021}) and has been accepted by the community as a credible way to abstract large scale networked systems that are already quite complex to model.

The transfer of energy between the nodes is impaired by the loss of energy in the process, and subsequently, the receiving node can only receive a fraction of the transmitted energy. For example, if $e$ energy is transferred from node $u_i$ to $u_j$ over a time duration ($t$,$t'$), and the initial energy levels of these nodes at time $t$ were $E_t'(i)$ and $E_t'(j)$, the remaining energy of these nodes at time $t'$ will be, 
\begin{equation}
\big( E_t'(i), E_t'(j) \big) = \big( E_t(i)-e, E_t(j)+(1-\beta) e \big)
\label{eq:peer_rem_energy},
\end{equation}
where the energy loss factor is denoted by $\beta \in [0,1)$. We also assume that $\beta$ remains constant during the whole WPT process, as typically in the related literature, such as \cite{Raptis2020} \cite{Bulut2020social}. In addition, the energy transfer between any two nodes $u_i$ and $u_j$ does not have any effect on the energy levels of other nodes ($\forall u_k \in \mathcal{U}, u_k \ne u_i, u_j$), and is also mutually exclusive of the energy transfers between any other pairs of nodes ($u_k$ and $u_l$, $\forall u_k, u_l \ne u_i, u_j$). 

Further, we define the energy distribution of the deployed nodes ($\mathcal{E}_t(u)$) at any given time $t$, over the sample space of set of nodes $\mathcal{U}$ as,
\begin{equation}
\mathcal{E}_t(u) = \frac{E_t(u)}{E_t(\mathcal{U})} \qquad \mbox{ where, } E_t(\mathcal{U}) = \sum_{u \in \mathcal{U}} E_t(u)
\end{equation}
Subsequently, the average network energy can be computed as,
\begin{equation}
\overline{E}_t = \frac{E_t(\mathcal{U})}{m}.
\end{equation}

The parameter \textit{energy variation distance} among the deployed nodes provides an overall estimate of the energy distribution of the whole network. To compute the variation distance, we use the probability theory and stochastic processes defined as described by \cite{Nikoletseas2017, Dhungana2021}. We consider, two probability distributions, namely $P$ and $Q$, defined over the sample space of $\mathcal{U}$. Now, the total variation distance, $\delta(P,Q)$, is computed using Equation \ref{eq:var_dist}.
\begin{equation}
\delta(P,Q) = \sum_{x \in \mathcal{U}} |P(x) - Q(x)|
\label{eq:var_dist}
\end{equation}

In our model, the devices are carried by users and are therefore characterized by a human like mobility pattern. We assume that the users move from one location to another following their own interest. Therefore, the users' movement for visiting different places depends on diverse factors, such as the interest to visit the place as well as the number of social connections in that place. In Section \ref{sec:Mobi_Soc_WEB}, we mathematically define the user movement and subsequently discuss the proposed method.

Thus, the number of users at different location vary over time due to the movement of users from one location to another. We use the following definition to compute this,
\begin{definition}
The number of users at any location $l$ at a given time $t$ is defined by $\phi_t(l)$. It is computed as the number of users $u_i \in \mathcal{U}$ for which $l = l^i_t$ in $L(t)$,
\begin{equation}
\phi_t(l) = |\mathtt{L}(t)|_{l = l^i_t}	\qquad \forall u_i \in \mathcal{U}
\end{equation}
\end{definition}

In our model, we assume that users meet one another when they remain at the same location at the same time and also, spend a certain duration of time. To perform P2P-WPT between two users, the specific conditions for valid contact between the users need to be satisfied.  
\begin{definition}
A valid contact ($\nu_{ij}^t$) is defined as the contact between any two nodes $u_i$ and $u_j$ for a duration of $(t',t'')$ satisfying the following conditions $\forall t \in (t',t'')$,
\begin{equation} 
\nu_{ij}^t = 
\begin{cases}
1, \quad (\overline{l^i_t, l^j_t}) \leq d_{req} \mbox{ and } (t'' - t') \geq t_{min},\\
0, \quad \mbox{otherwise}
\end{cases}
\end{equation}
where we represent the distance between the locations of $u_i$ and $u_j$ by $(\overline{l^i_t, l^j_t})$ and the required allowable distance for performing WPT is denoted by $d_{req}$. $t_{min}$ refers to the minimum time required for performing successful P2P-WPT energy transfer.
\end{definition}

All the symbols used in our proposed method are listed in Table \ref{table:symbols}.
\begin{table*}
\centering
\caption{List of Symbols}
\begin{tabular}{|l|p{8cm}|}
\hline
\textbf{Symbol} & \textbf{Meaning}\\ 
\hline
$m$			& The number of users/mobile devices \\ \hline
$\mathcal{U}$	& The set of users \\ \hline
$n$			& The number of locations \\ \hline
$\mathbb{L}$	& The set of the localized nodes \\ \hline
$l^i_t$			& Location of any user $u_i$ at time $t$ \\ \hline
$\mathtt{L}(t)$	& The set of the locations of all the users at time $t$ \\ \hline
$\mathcal{E}_t$	& The set of energy levels of the nodes at time $t$ \\ \hline
$\beta$			& The energy loss factor \\ \hline
$\alpha$		& The rate of energy transfer per unit time \\ \hline
$E_t(u)$		& Total network energy at time $t$ \\ \hline
$\overline{E}_t$ & Average network energy at time $t$ \\ \hline
$\delta(P,Q)$	& Total variation distance for two probability distributions, $P$ and $Q$\\ \hline
$\phi_t(l)$		& The number of users at any location $l$ at a given time $t$ \\ \hline
$\nu_{ij}^t$	& Valid contact between any two users $u_i$ and $u_j$ \\ \hline
$\mathtt{H}^i_t$ & The mobility history of user $u_i$ till time $t$ \\ \hline
$\mathtt{P}^i_t$ & The set of transition probabilities of user $u_i$ till time $t$ \\ \hline
$\mathtt{L}_i(t)$ & The sequence of locations visited by user $u_i$ till time $t$ \\ \hline
$\mathtt{Z}_i(t)$ & The set of arrival times at $\mathtt{L}_i(t)$ \\ \hline
$\mathtt{S}_i(t)$ & The set of duration of stay at $\mathtt{L}_i(t)$ \\ \hline
$c_i$			& The location context of user $u_i$ \\ \hline
$N(c_i,\mathtt{H}^i_t)$ & The number of times the pattern $c_i$ appear in $\mathtt{H}^i_t$ \\ \hline
$State[i]$		& Current state of user $u_i$ \\ \hline
$\lambda_t(u_i)$ & The time elapsed for user $u_i$ \\ \hline
$\tau^{i,l}_t$	& Average stay duration of user $u_i$ in any location $l$ \\ \hline
$\mathcal{LA}^{u_i,l}_t$ & Location attachment of user $u_i$ towards location $l$ \\ \hline
$\mathcal{PS}_{sc}(u_i, u_j)$ & PeerSelectivity based on social contexts between users $u_i$ and $u_j$ \\ \hline
$\mathcal{SC}(u_i, u_j)$ & Social Connection between users $u_i$ and $u_j$ \\ \hline
$\mathcal{SA}^{u_i}_t$   & Social Attachment of user $u_i$ at time $t$ \\ \hline
$\mathcal{PS}_{scr}(u_i, u_j)$ & PeerSelectivity based on social context and relations between users $u_i$ and $u_j$ \\ \hline
$w_l$, $w_s$, $w_e$ & The weightage for the location attachment, social attachment and energy components \\ \hline
\end{tabular}
\label{table:symbols}
\end{table*}

\subsection{MoSaBa Main Concept}

\begin{figure*}[!ht]
\centering
\includegraphics[scale=0.5]{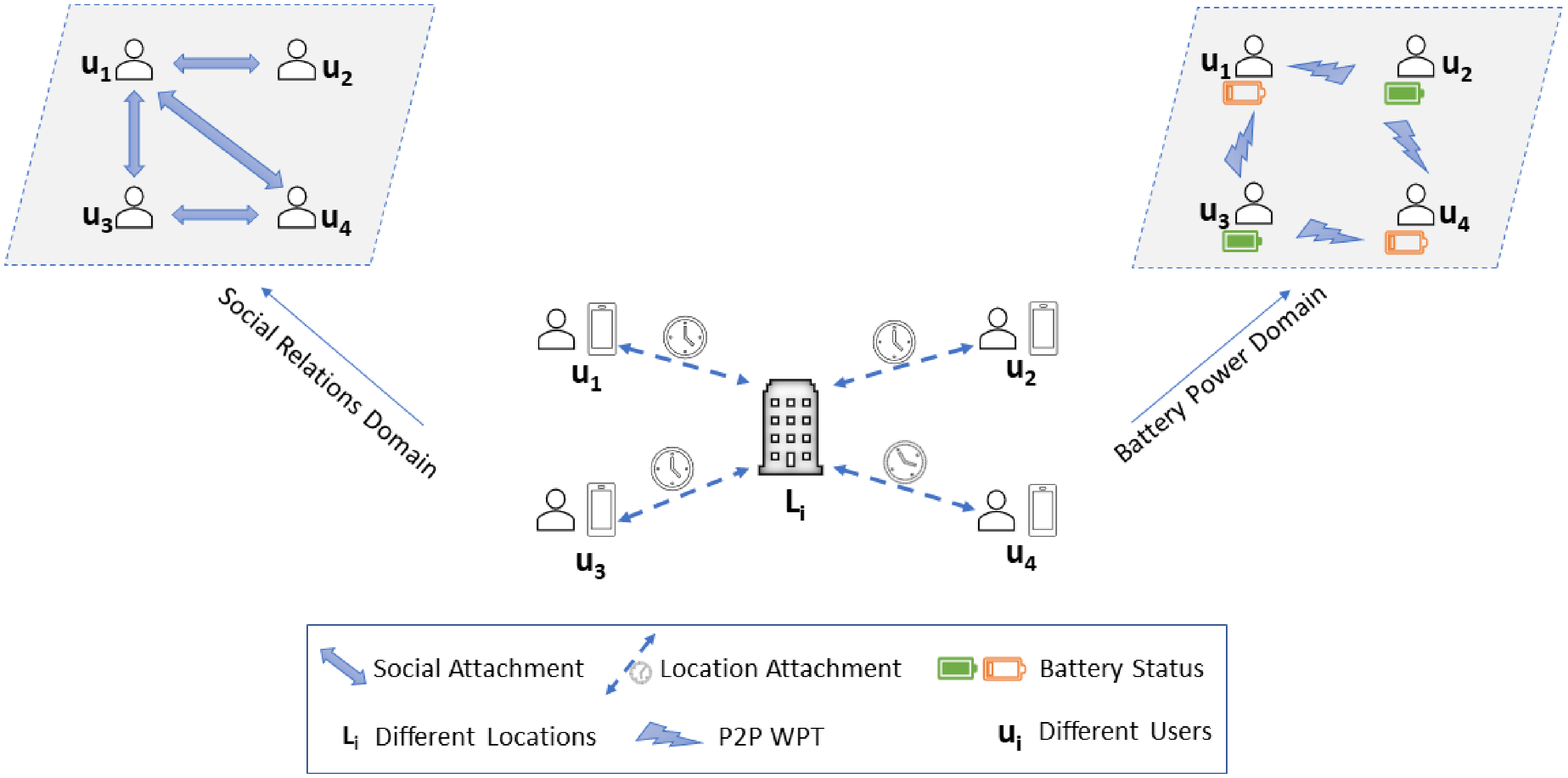}
\caption{The network model with an example scenario}
\label{fig:net_model}
\end{figure*}

In Figure \ref{fig:net_model}, we present a diagram of the network model with an example scenario. This figure shows one of the locations of the whole scenario area having different number of users present in that location at any specific time. We show that each user has `location attachment', which determines the expected time before the user can move to another location. The social relations among these users are depicted in the `social relations domain'. Similarly, the `battery power domain' shows the current remaining energy status of these users. The users present at the same location can engage in P2P-WPT with the other users (transmits or receives energy depending on their battery level). Based on all these information, in the proposed method, the peers are selected for P2P-WPT to reach energy balance throughout the crowd. For example, as shown in Figure \ref{fig:net_model}, the location $L_i$ has four different users. These four users have different location attachment values and different remaining energy. Also, social contacts between these users are known. From the figure, it is evident that $u_2$ and $u_3$ can transmit energy to other two nodes, i.e., $u_1$ and $u_4$, because they have suitable energy levels, and due to their social relationships, we can predict that they will remain co-located for sufficient amount of time. Now, based on all these information, our proposed method will select the peers such that the crowd energy is balanced with low energy loss and high remaining energy. The other locations of the scenario also goes through same situation. The users move from one location to another based on the influence of social/location attachment.

\section{Mobility and Social-aware Wireless Energy Balancing}\label{sec:Mobi_Soc_WEB}
This section presents the proposed \textit{MoSaBa} method in detail. First, in Section \ref{sec:sec:mob_pred}, we discuss the prediction of user mobility. Then, in Section \ref{sec:sec:soc_cont}, we discuss social contexts for the users, and compute the impact of social contexts and user mobility in peer selection for WPT. Next, in Section \ref{sec:sec:soc_cont_rel}, we compute and discuss the influence of social relations on the peer selection process. Subsequently, we present the mobility and social-aware energy balancing process adopted in MoSaBa. For each part, we present an algorithm enlisting the steps followed in the proposed method.

\subsection{User mobility prediction}\label{sec:sec:mob_pred}
We employ the $O(k)$ Markov predictor \cite{Song2006, Chon2014} to predict the future locations of the users. It is considered as one of the promising prediction-based approaches for human mobility prediction, and also for prediction based on past location history \cite{Song2006a}. Here, the $O(k)$ Markov predictor considers the mobility history of the users and the recent $k$ locations, which is called the \textit{location context}, to predict the future locations. For each user, the corresponding location history is searched for the \textit{location context}, and subsequently estimate the probabilities for moving to any location next.  

The mobility prediction model needs two sets of information -- the mobility history ($\mathtt{H}^i_t$) and the set of transition probabilities ($\mathtt{P}^i_t$). The mobility history $\mathtt{H}^i_t$ consists of three sets of information -- the sequence of locations visited till time $t$ by user $u_i$ ($\mathtt{L}_i(t) = \{l^i_1, l^i_2, \cdots, l^i_t\}$), the set of arrival times at these locations ($\mathtt{Z}_i(t) = \{z^i_1, z^i_2, \cdots, z^i_t\}$), and the set of duration of stay at these locations ($\mathtt{S}_i(t) = \{s^i_1, s^i_2, \cdots, s^i_t\}$). The transition probability $p^i_{ab} \in \mathtt{P}^i_t$ refers to the probability of moving from location $l^i_a$ to location $l^i_b$ for any user $u_i$. The values of the transition probabilities are computed from the information stored in the location history set. This process is repeated for all the users with their corresponding information.

It is noteworthy to mention that the P2P-WPT method requires information for both locations as well as the meeting duration between the users to determine the suitable pairs of peers. Therefore, at any given time $t$, the model computes the transition probability ($P_i(l^i_{t+1} = x|\mathtt{H}^i_t)$) for different locations ($x \in \mathbb{L}$) as well as an estimate of the stay duration -- i.e. whether the user can move to any other location within the next $\Delta t$ time.

We first compute the location context $c_i$ of user $u_i$ by extracting the sequence of recently visited $k$ locations from $\mathtt{L}_i(t)$. Thus, $c_i = \mathtt{L}_i(t-k+1, t) = \{l^i_{t-k+1}, l^i_{t-k+2}, \cdots, l^i_{t-1}, l^i_t\}$. As mentioned previously, to compute the transition probabilities and estimate of stay duration, the mobility predictor uses the location context information to find matching patterns in the location history $\mathtt{L}_i(t)$. Next, we compute the set of duration for any such possible location $x$ using the following,
\begin{equation}
S^i_x = \{ s^i_t|s^i_t=z^i_{t+1}-z^i_t \mbox{ where } \mathtt{L}_i(t-k+1, t+1)=c_i x \}
\label{eq:S_x}
\end{equation}
where $c_i x$ is the sequence of locations assuming the location $x$ will be visited after $c_i$. Next, using the duration set $S^i_x$, the conditional probability $P^i_x(t \leq s < t+\Delta t | c_i,t )$ for the user to move to location $x$ within $\Delta t$ time period of the current elapsed time $t$. Here, we use the CDF predictor described in \cite{Chon2014}, and subsequently we can compute the following:
\begin{align}
P^i_x(t \leq s < t+\Delta t | c,t ) \nonumber \\
&\hspace*{-9em}\approx CDF(s < t + \Delta t) - CDF(s < t) \nonumber \\
&\hspace*{-9em}= \frac{1}{|S^i_x|}\sum_{s \in S^i_x} I(s < t + \Delta t) - \frac{1}{|S^i_x|}\sum_{s \in S^i_x} I(s < t)
\end{align}
where $I(\cdot)$ refers to the indicator function.

We compute the transition probability for any possible location $x \in \mathbb{L}$ as,
\begin{equation}
P_i(l^i_{t+1} = x|\mathtt{H}^i_t) \approx \widehat{P}(l^i_{t+1} = x|\mathtt{H}^i_t) = \frac{N(c_i x,\mathtt{H}^i_t)}{N(c_i,\mathtt{H}^i_t)}
\end{equation}
Here, $N(c_i x,\mathtt{H}^i_t)$ and $N(c_i,\mathtt{H}^i_t)$ denote the number of times the pattern $c_i x$ and $c_i$, respectively, appear in $\mathtt{H}^i_t$. 

It is important to note that the $O(k)$ Markov predictor will fail to provide the predictions in case the current location context has not appeared till $(t-1)$ time (i.e., $N(c_i,\mathtt{H}^i_{t-1}) = 0$). In such situation, the $O(k-1)$ predictor is applied next for finding the location predictions, and the process is followed again in case of the model being unable to predict. In case all previous estimations are failed, the `order-0' predictor computes the most frequently visited location for the user, i.e., $l = \argmax_{l\in \mathtt{L}_i(t)} N(l,\mathtt{H}^i_t)$.  

Next, we estimate whether the user can move to any location $x$ within $\Delta t$ time. With the information on the given location context $c_i$ and time $t$, we can compute that,
\begin{equation}
P_i(x|c_i,t) = P_i(x)\times P^i_x(t \leq s < t+\Delta t | c_i,t )
\label{eq:prob_loc_x}
\end{equation}

Subsequently, we can compute the most likely next location for time $t+1$,
\begin{equation}
l^i_{t+1} = \argmax_{x \in \mathtt{L}_i(t)} P(l^i_{t+1}=x|c_i,t)
\label{eq:prob_nex_loc}
\end{equation} 
Similarly, the mobility predictor estimates the probable locations for each mobile node $u_i \in \mathcal{U}$.

In the following, we discuss the process followed by the nodes for performing P2P-WPT, based on the already computed information. The objective of the P2P-WPT method is to an energy balance of the whole network such that the energy loss   and the energy variation distance is minimized. Ideally, at energy balance, each node of the network should attain the average network energy ($\overline{E}_t$) either by transferring or receiving energy. However, as the WPT process is impaired by the loss of energy during the process, the nodes can not practically reach $\overline{E}_t$, rather can achieve $\overline{E}^*$. As shown by \cite{Dhungana2019}, the computation of average network energy with energy loss is as follows, 
\begin{equation}
\overline{E}^* = \frac{-(1-\beta) + \sqrt[2]{(1-\beta)}}{\beta} \qquad \forall \beta \in [0, 1], m \longrightarrow \infty
\end{equation}
\cite{Dhungana2019} showed that $\overline{E}^* \in [0,1] $ for $\beta \in [0,1]$. It is important to note that if we select the users with energy at the opposite side of $\overline{E}^*$ for engaging in P2P-WPT, the energy loss in the process will be reduced. Additionally, the users having energy closer to $\overline{E}^*$ should be selected for exchanging energy first, such that the decrease of energy variation distance in each iteration will be higher compared to other combinations.  

In our scenario, the mobility of the users impacts the number of users they can meet as well as the duration of such meetings. Therefore, it is important to carefully choose the peers in each iteration. Algorithm \ref{algo:mobi_energy_balance} describes the steps for mobility-aware energy balancing. As discussed, to minimize the energy loss and energy variation distance, we choose a pair of nodes ($u_i, u_j$) having their current energy values at the opposite side of $\overline{E}^*$, and also, their current energy levels are closest to the target energy balance level ($\overline{E}^*$). Next, for each of such chosen pairs of peers, P2P-WPT is performed according to the steps mentioned in the Algorithm \ref{algo:p2p_energy_balance}, as explained in Section \ref{sec:sec:p2p_eb}.

\begin{algorithm}[t!]
\caption{Mobility-aware Energy Balancing\label{algo:mobi_energy_balance}}
\textbf{Inputs:} Set of energy levels $\mathcal{E}_t$, Set of predicted location $\{l^i_t\}_{\forall u_i \in \mathcal{U}}$, Set of stay duration $\{S^i_x\}_{x \in \mathtt{L}_i(t)}^{t}$.\\
\textbf{Output:} Energy balance of the network.\\
\SetKwFunction{MyFunc}{MobiEnergyBalance}
Compute $\overline{E}^*$\;
Initialize $State[\cdot] \longleftarrow Incomplete$\;
\While{$t \leq T$}{
\For{$u_i \in \mathcal{U}$ and $State[i] = Incomplete$}{
	Find the node with energy closer to $\overline{E}^*$, $u_i \longleftarrow \argmin_{u_i \in \mathcal{U}} |\overline{E}^* - E_t(i)|$\;
	\texttt{MobiEnergyBalance}($\mathcal{E}_t$, $u_i$, $t$, $State[\cdot]$)\;
}
\For{$u_i \in \mathcal{U}$ and $State[i] = Incomplete$}{
	Compute $t' \longleftarrow \min_{u_i \in \mathcal{U}} \{ \lambda_t(u_i) \} $\;
	\If{$t' < t + \Delta t$}{
		$u'_i \longleftarrow \argmin_{u_i \in \mathcal{U}} \{ \lambda_t(u_i) \} $\;
		\texttt{MobiEnergyBalance}($\mathcal{E}_t'$, $u'_i$, $t'$, $State[\cdot]$)\;
	}
}
\If{Current time $= t + \Delta t$}{
	Get $\{l^i_{t + \Delta t}\}_{\forall u_i \in \mathcal{U}}$ and $\{S^i_x\}_{x\in \mathtt{L}_i(t)}^{t + \Delta t}$ from mobility predictor\;
	Update $t \longleftarrow t + \Delta t$\;
}
}
\SetKwProg{Fn}{Function}{:}{}
\Fn{\MyFunc{$\mathcal{E}_t$, $u_i$, $t$, $State[\cdot]$}}{
	\For{$u_j \in \mathcal{U}$ and $u_j \neq u_i$}{
		\If{$\nu_{ij}^t = 1$ }{ 
			$N_t(u_i) \longleftarrow N_t(u_i) \cup u_j$\;	
		}
	}
	\eIf{$E_t(i) > \overline{E}^*$} { 
		Compute $u_j \longleftarrow \argmin_{u_j \in N_t(u_i)} ( \overline{E}^* - E_t(j) )$\;
		\texttt{EnergyBalance}($\mathcal{E}_t$, $u_i$, $u_j$, $State[\cdot]$)\; 
	}{ 
		Compute $u_j \longleftarrow \argmin_{u_j \in N_t(u_i)} ( E_t(j) - \overline{E}^* )$\;
		\texttt{EnergyBalance}($\mathcal{E}_t$, $u_j$, $u_i$, $State[\cdot]$)\; 
	}
} 
\end{algorithm}


\subsection{Impact of Social Context on Peer Selection for P2P-WPT}\label{sec:sec:soc_cont}
In this section, we discuss the impact of the social context on peer selection for P2P-WPT. The social context considered in this work refers to user's interest in visiting different places depending on the category of the place or the things the place offers. We model the concept of social context in such way that the users' preferences are not required to be known a priori. 

Towards modeling the impact of social contexts, continuing from the previous sections, we define the following parameters. Our model utilizes the information that we already have in the following computations. We first compute the average stay duration, as shown in Definition \ref{def:avg_stay_dur}, for any user at any specific location. 

\begin{definition}
Average stay duration ($\tau^{i,l}_t$) defines the average time any user $u_i$ spends in any location $l$ till time $t$. It is computed as using the following equation,
\begin{equation}
\tau^{i,l}_t = \frac{\sum_{t} s^i_t}{N(l, \mathtt{H}^i_t)}		\qquad s^i_t \in S^i_l, u_i \in \mathcal{U}
\end{equation} 
Here, $N(l, \mathtt{H}^i_t)$ denotes the number of times $u_i$ has visited location $l$. 
\label{def:avg_stay_dur}
\end{definition}

Next, we compute the \textit{location attachment} for any user towards any location. This parameter helps us in conceptualizing the impact of location-based social context in users' mobility. As mentioned previously, from the P2P-WPT point of view, the estimation of the meeting times between pairs of users is expected to be beneficial. Based on the information of location attachment, subsequently, we compute the parameter called \textit{PeerSelectivity}, which quantifies the benefit of selecting this pair of nodes towards reaching the goal of energy balance. We explain this parameter in Definition \ref{def:ps_sc}. In Algorithm \ref{algo:social_context_energy_balance}, we provide the steps followed for social context-aware peer selection towards energy balancing in P2P-WPT. 

\begin{definition}
Location attachment ($\mathcal{LA}^{u_i,l}_t$) refers to the interest of any user $u_i$ towards staying at location $l$ based on his/her interests matching with that of this location's. This is computed as the fraction of total time the user spends in any location with respect to the total time spent in different places visited by this user. Mathematically,
\begin{equation}
\mathcal{LA}^{u_i,l}_t = \frac{\tau^{i,l}_t}{\sum_{l \in \mathtt{H}^i} \tau^{i,l}_t}		\qquad u_i \in \mathcal{U}
\end{equation} 
\end{definition}

\begin{definition}
PeerSelectivity ($\mathcal{PS}_{sc}(u_i, u_j)$) defines the selectivity factor between the peers $u_i$ and $u_j$ based on their social contexts (the first part of the Eq. \eqref{eq:ps_eqn} refers to the social context dimension). In other words, this parameter quantifies the benefit of selecting this pair of nodes towards reaching the goal of energy balance. The lower the value of selectivity factor, the better the corresponding pair is for the crowd energy balance. It is computed as,
\begin{equation}
\mathcal{PS}_{sc}(u_i, u_j) = w_l \times \big(|\mathcal{LA}^{u_i,l}_t - \mathcal{LA}^{u_j,l}_t | \big) + w_e \times \big( \frac{E_1 - E_2}{E_{max}} \big) \label{eq:ps_eqn}
\end{equation}
where $E_1$ and $E_2$ refers to required energy levels for energy balancing. For example, when $E_t(i) > \overline{E}^*$, $E_1 = \overline{E}^*$ and $E_2 = E_t(j)$. Similarly, for $E_t(i) < \overline{E}^*$, the values are opposite, i.e., $E_1 = E_t(j)$ and $E_2 = \overline{E}^*$. $E_{max}$ refers to the maximum possible energy level of any user, and thus, $0 \leq E_1, E_2 \leq E_{max}$. Here, $w_l$ and $w_e$ denote the weightage for the location attachment and energy components, respectively.
\label{def:ps_sc}
\end{definition}

\begin{algorithm}[t!]
\caption{Social Context-aware Energy Balancing\label{algo:social_context_energy_balance}}
\textbf{Inputs:} Set of energy levels $\mathcal{E}_t$, Set of stay duration $\{S^i_x\}_{x \in \mathtt{L}_i(t)}^{t}$.\\
\textbf{Output:} Energy balance of the network.\\
\SetKwFunction{MyFunc}{SocialContext}
\SetKwProg{Fn}{Function}{:}{}
\Fn{\MyFunc{$\mathcal{E}_t$, $u_i$, $t$, $State[\cdot]$}}{
	\For{$u_j \in \mathcal{U}$ and $u_j \neq u_i$}{
		\If{$\nu_{ij}^t = 1$ }{ 
			$N_t(u_i) \longleftarrow N_t(u_i) \cup u_j$\;	
		}
	}
	Compute $\tau^{i,l}_t$ and $\mathcal{LA}^{u_i,l}_t$ for $u_i$\;
	\For{$u_j \in \mathcal{U}$ and $u_j \neq u_i$}{
		Compute $\tau^{j,l}_t$ and $\mathcal{LA}^{u_j,l}_t$ for $u_j$\;	
	}
	\eIf{$E_t(i) > \overline{E}^*$} { 
		Set $E_1 \longleftarrow \overline{E}^*$, $E_2 \longleftarrow E_t(j)$\;
		Compute $u_j \longleftarrow \argmin_{u_j \in N_t(u_i)} \mathcal{PS}_{sc}(u_i, u_j) $\;
		\texttt{EnergyBalance}($\mathcal{E}_t$, $u_i$, $u_j$, $State[\cdot]$)\; 
	}{ 
		Set $E_1 \longleftarrow E_t(j)$, $E_2 \longleftarrow \overline{E}^*$\;
		Compute $u_j \longleftarrow \argmin_{u_j \in N_t(u_i)} \mathcal{PS}_{sc}(u_i, u_j) $\;
		\texttt{EnergyBalance}($\mathcal{E}_t$, $u_j$, $u_i$, $State[\cdot]$)\; 
	}
} 
\end{algorithm}

To motivate the usage of the `social context' dimension, as well as to demonstrate its effectiveness in the P2P-WPT process, we evaluate the performance of the social context-aware algorithm over the mobility-aware algorithm using simulations. Please refer to Section \ref{sec:sec:eff_sc} for the detailed results.

\subsection{Impact of Social Relations on Peer Selection for P2P-WPT}\label{sec:sec:soc_cont_rel}
In this section, we present the concepts for modeling the impact of social relations on peer selection for P2P-WPT. Continuing from the previous section, here, we introduce few more parameters to quantify the impact of users' social relation on peer selection. First, we maintain a parameter named Social Connection ($\mathcal{SC}(u_i, u_j)$) to note the social connections between the users $u_i$ and $u_j$. Next, we compute the social attachment for any user depending on the number of social connections the user has in its current location in that specific time. It is important to note that this information needs to be computed at each iterations, unlike the location attachment, as the value changes according to the location as well as other users present in that location.

\begin{definition}
Social Connection ($\mathcal{SC}(u_i, u_j)$) refers to the existence of any social connection, i.e., holds the value of the edge $(u_i, u_j)$ on the social graph between these users. In practice, this information can be computed from the users' self reported social network data. It is mathematically defined as,
\begin{equation} 
\mathcal{SC}(u_i, u_j) = 
\begin{cases}
1, \quad \mbox{Social edge exists} (j \ne i),\\
0, \quad \mbox{otherwise}
\end{cases}
\end{equation}
\end{definition}

\begin{definition}
Social Attachment ($\mathcal{SA}^{u_i}_t$) defines the overall social attachment any user $u_i$ has with the other users ($\forall u_j \in \mathcal{U}, j \ne i$) at time $t$. It is computed as the number of users socially connected with $u_i$ among all the users present at the same location at time $t$.
\begin{equation}
    \mathcal{SA}^i_t = \frac{\sum_{\forall u_j \in \mathcal{U}, j \ne i} \mathcal{SC}(u_i,u_j)}{\phi_t(l)}
\end{equation}
\end{definition}

\begin{definition}
PeerSelectivity based on social context and relations ($\mathcal{PS}_{scr}(u_i, u_j)$) defines the selectivity factor between the peers $u_i$ and $u_j$ based on their social contexts. It is computed as,
\begin{multline}
\mathcal{PS}_{scr}(u_i, u_j) = \mathcal{PS}_{sc}(u_i, u_j) +  w_s \times \\ \big(|\mathcal{SA}^{u_i,l}_t - \mathcal{SA}^{u_j,l}_t | \big)
\label{eq:ps_scr}
\end{multline} 
where $w_s$ refers to the weightage for the social attachment in the P2P-WPT process.
\end{definition}

Based on the information of social connections and the dynamically computed social attachment, we compute the selectivity factor for any pair of peers. The selectivity factor also considers both the location and social attachment. Algorithm \ref{algo:social_context_relation_energy_balance} lists the steps followed for the social context and relation-aware peer selection for P2P-WPT.

\begin{algorithm}[t!]
\caption{Social Context and Relations-aware Energy Balancing\label{algo:social_context_relation_energy_balance}}
\textbf{Inputs:} Set of energy levels $\mathcal{E}_t$, Set of stay duration $\{S^i_x\}_{x \in \mathtt{L}_i(t)}^{t}$.\\
\textbf{Output:} Energy balance of the network.\\
\SetKwFunction{MyFunc}{SocialContextRelations}
\SetKwProg{Fn}{Function}{:}{}
\Fn{\MyFunc{$\mathcal{E}_t$, $u_i$, $t$, $State[\cdot]$}}{
	\For{$u_j \in \mathcal{U}$ and $u_j \neq u_i$}{
		\If{$\nu_{ij}^t = 1$ }{ 
			$N_t(u_i) \longleftarrow N_t(u_i) \cup u_j$\;	
		}
	}
	Compute $\tau^{i,l}_t$ and $\mathcal{LA}^{u_i,l}_t$ for $u_i$\;
	Compute $\mathcal{SC}(u_i, u_j)$ and $\mathcal{SA}^{u_i}_t$ for $u_i$\;
	\For{$u_j \in \mathcal{U}$ and $u_j \neq u_i$}{
		Compute $\tau^{j,l}_t$ and $\mathcal{LA}^{u_j,l}_t$ for $u_j$\;	
		Compute $\mathcal{SC}(u_i, u_j)$ and $\mathcal{SA}^{u_i}_t$ for $u_j$\;
	}
	\eIf{$E_t(i) > \overline{E}^*$} { 
		Set $E_1 \longleftarrow \overline{E}^*$, $E_2 \longleftarrow E_t(j)$\;
		Compute $u_j \longleftarrow \argmin_{u_j \in N_t(u_i)} \mathcal{PS}_{scr}(u_i, u_j) $\;
		\texttt{EnergyBalance}($\mathcal{E}_t$, $u_i$, $u_j$, $State[\cdot]$)\; 
	}{ 
		Set $E_1 \longleftarrow E_t(j)$, $E_2 \longleftarrow \overline{E}^*$\;
		Compute $u_j \longleftarrow \argmin_{u_j \in N_t(u_i)} \mathcal{PS}_{scr}(u_i, u_j) $\;
		\texttt{EnergyBalance}($\mathcal{E}_t$, $u_j$, $u_i$, $State[\cdot]$)\; 
	}
} 
\end{algorithm}

To motivate the effect of social relations in P2P-WPT, we evaluate and compare the performance of the social context and relation-aware peer selection method. Please refer to Section \ref{sec:sec:eff_scr} for the detailed evaluation results.

\subsection{P2P Energy Balancing}\label{sec:sec:p2p_eb}
In Algorithm \ref{algo:p2p_energy_balance}, we discuss the steps followed for P2P-WPT between any selected pair of peers. Here, among the pair, the user which is having energy closer to the $\overline{E}^*$ (considering the energy loss in the process), is allowed to reach the target energy balance level (i.e, $\overline{E}^*$) first. This energy exchange is bounded by the energy transfer limit ($\alpha.t_{p2p}$), which is the possible value of energy that can be transferred within the duration $t_{p2p}$ with $\alpha$ rate of energy transfer per unit time. Here, $t_{p2p}$ is computed as minimum of stay durations of the nodes and $\Delta t$ -- the actual duration of time for which the energy exchange can be possible. Based on these information, the energy levels of the users are updated, and the time elapsed for the participating users ($\lambda_t(u_i), \lambda_t(u_i)$) are also updated. During the process, the state variable $State[\cdot]$ keeps track of the current state, \textit{Incomplete}, \textit{Busy}, or \textit{Complete}, of the users.  For example, the state of the users remains \textit{Busy} during the P2P energy exchange process. The users which are yet to reach the target energy balance have \textit{Incomplete} state, and those who have achieved energy balance are marked as \textit{Complete}. Here, the users with state \textit{Incomplete} but with elapsed time within the current time iteration's duration, i.e., $\lambda_t(u_i) < \Delta t$, are again considered as candidates for possible energy balance. Then, when the current iteration ends, i.e., $t + \Delta t$, these steps are repeated for each iteration till the final time $T$.

\begin{algorithm}[t!]
\caption{P2P Energy Balance\label{algo:p2p_energy_balance}}
\textbf{Inputs:} Set of energy levels $\mathcal{E}_t$, Nodes $u_1$ and $u_2$, Set of flags $State[\cdot]$.\\
\textbf{Output:} Updated energy levels and state, time elapsed.\\
\SetKwFunction{MyFunc}{EnergyBalance}
\SetKwProg{Fn}{Function}{:}{}
\Fn{\MyFunc{$\mathcal{E}_t$, $u_1$, $u_2$, $State[\cdot]$}}{
    Get stay duration for $u_1$ and $u_2$ as $s_1$ and $s_2$\;
	Compute effective P2P meeting time, $t_{p2p} \longleftarrow min_{} (s_1, s_2, \Delta t)$\;
	\eIf{$[E_t(u_1) - \overline{E}^*](1-\beta) < [\overline{E}^* - E_t(u_2)]$}{ 
		Check whether required energy transfer is within effective meeting time, $\eta \longleftarrow ( |E_t(u_1) - \overline{E}^*| \leq \alpha.t_{p2p} )? 0:1$\;
		Update the energy levels considering the energy transfer limit,
		$E_t(u_1) \longleftarrow \overline{E}^* + \eta[E_t(u_1) - \overline{E}^* - \alpha.t_{p2p}]$\;
		$E_t(u_2) \longleftarrow E_t(u_2) + [ E_t(u_1) - \overline{E}^* - \eta(E_t(u_1) - \overline{E}^* - \alpha.t_{p2p}) ](1-\beta) $\;		
		Time required for P2P energy exchange, $\lambda \longleftarrow \frac{1}{\alpha}[ E_t(u_1) - \overline{E}^* - \eta(E_t(u_1) - \overline{E}^* - \alpha.t_{p2p}) ]$\;
		\If{Current time $< t+\lambda$}{
			Set $State[u_1] \longleftarrow Busy$, $State[u_2] \longleftarrow Busy$\;	
		}
		\If{$\eta = 0$ and Current time $= t+\lambda$}{
			Set $State[u_1] \longleftarrow Complete$, $State[u_2] \longleftarrow Incomplete$\;	
		}
	}{ 
		Check whether required energy transfer is within effective meeting time, $\eta \longleftarrow ( |\overline{E}^* - E_t(u_2)| \leq \alpha.t_{p2p} )? 0:1$\;
		Update the energy levels considering the energy transfer limit,
		$E_t(u_1) \longleftarrow E_t(u_1) - \frac{ [\overline{E}^* - E_t(u_2)] - \eta [\overline{E}^* - E_t(u_2) - \alpha.t_{p2p}] }{ (1-\beta) }$\;
		$E_t(u_2) \longleftarrow \overline{E}^* - \eta[\overline{E}^* - E_t(u_2) - \alpha.t_{p2p}]$ \;
		Time required for P2P energy exchange, $\lambda \longleftarrow \frac{1}{\alpha}[\overline{E}^* - E_t(u_2)] + \eta [\alpha.t_{p2p} + \overline{E}^* - E_t(u_2)]$\;		
		\If{Current time $< t+\lambda$}{
			Set $State[u_1] \longleftarrow Busy$, $State[u_2] \longleftarrow Busy$\;	
		}
		\If{$\eta = 0$ and Current time $= t+\lambda$}{
			Set $State[u_1] \longleftarrow Incomplete$, $State[u_2] \longleftarrow Complete$\;	
		}
	}
	Update $\lambda_t(u_1) \longleftarrow \lambda$, $\lambda_t(u_2) \longleftarrow \lambda$\; 
} 
\end{algorithm}

\subsection{Complexity of the Algorithms}
In this section, we explain the computational complexity of the proposed algorithms and its theoretical analysis. The Algorithms \ref{algo:mobi_energy_balance}, \ref{algo:social_context_energy_balance}, and \ref{algo:social_context_relation_energy_balance} have an execution time in $O(m^3)$, where $m$ denotes the number of users. Here, in all three algorithms, the peer selection process has an execution time in $O(m^2)$ as it requires $m-1$ comparisons to find the first peer, and then $m-2$ comparisons to find the second peer for each first peer. Now, peer selection is performed for $m$ number of nodes. Therefore, the whole process has an execution time in $O(m^3)$.

On the other hand, the Algorithm \ref{algo:p2p_energy_balance} has a constant execution time ($O(1)$), as it only updates the energy levels of the two selected nodes each time. Whereas, the execution time for mobility prediction algorithm is in $O(T)$, where $T$ denotes the number of iterations.

\section{Performance Evaluation of the Social-related Dimensions of MoSaBa}\label{sec:Result_1}

\subsection{Simulation Settings}\label{sec:sim_params}
For performance evaluation, our experiments are based on an artificially generated scenario and not explicitly a dataset setting. We consider a crowd of 100 users each carrying a mobile device equipped with P2P-WPT facilities. The overall area has 5 different locations, where these 100 users are randomly distributed. In some of the experiments, we have considered 125 and 150 users present over the same number of locations. The stay duration in each location is influenced by user's interests and social interactions. The users spend a random time distributed over 10-40 $minutes$ in each location before moving to the next randomly selected location. Additionally, with the influence of the socially connected users, any user spends a random time in each location randomly distributed over $[0, f]$, where $f$ is the number of socially connected users at that location at that time. The inter-meeting duration for any pair of users is therefore random. It is important to note that in this simulation, both the location attachment and social attachment values are incorporated randomly. However, we can also preset these values for the users based on their preferences and self-reported social profiles. The users can engage in P2P-WPT with other users during their stay at the same location. The initial energy of the users are randomly distributed over $[0, 100]$ units. We consider the wireless charging rate of $\alpha = 0.5$ which closely mimics a real Qi charger\footnote{https://www.belkin.com/th/chargers/wireless/charging-pads/boost-up-wireless-charging-pad-7-5w/p/p-f7u027/} (a Qi charger with capacity of 7.5 $Wh$ will have $\alpha \approx 0.675$. In our experiments, we consider the energy loss rate $\beta = 0.2 - 0.4$. We choose $w_l = w_s = w_e = 0.33$ to emphasize equal weightage for all three components shown in Equation \eqref{eq:ps_scr}. 

We design our experiments to be performed over multiple `rounds' or `'iterations'. For example, in each iteration of the experiment, the users move to a randomly chosen location, stay for the mentioned random duration, and engage in P2P-WPT with their chosen peers. Therefore, an iteration actually refers to a virtual boundary within which the scope of decisions (about P2P meetings) remain. In all the experiments, we execute the concept of iterations, and accordingly, the decisions on finding peers are taken. It is noteworthy to mention that energy exchange between any two peers is limited by their meeting duration. Therefore, the concept of time is well enforced within the experiments. We perform the simulations using a laptop with an i5-1135G7 processor with possible CPU speed of $2.4-4.2$ $GHz$ and 16 GB of RAM. During the simulations, the CPU speed remains within $2.4-2.6$ $GHz$ and utilization remains $< 12\%$. We repeat the experiments 50 times and consider the average values for statistical smoothness.

\subsection{Evaluation Metrics}\label{sec:sec:eval_metrics}

\begin{itemize}
\item \textit{Total network energy}: Total network energy refers to the total network energy of all the nodes in the experiment at each iteration. This metric computes the remaining energy of the network in each iteration and helps us understand the energy loss occurring in each method. It is important to note that energy loss also increases with increased amount of energy exchanged.

\item \textit{Total energy variation distance}: The total energy variation distance measures the energy variation distance resulting between the nodes at each iteration. Using this metric, we can understand the quality of the energy balance -- the lower the number the better the quality of energy balance.

\item \textit{Number of P2P meetings}: This metric refers to the number of times a pair of peers engage in energy exchange. Using this metric, we can understand the number of P2P opportunities enabled by each method. 

\item \textit{Number of nodes that reach energy balance}: This metric counts the number of nodes which achieves energy balance or the expected energy target level. 

\item \textit{Execution time}: This metrics shows the actual CPU execution time in each iteration of the simulation for any method.

\end{itemize}

\subsection{Effect of Social Context in P2P-WPT}\label{sec:sec:eff_sc}
In order to motivate the usage of the `social context' dimension in the P2P-WPT process, as well as to demonstrate its effectiveness, we evaluate the performance of the social context-aware algorithm over the mobility-aware algorithm using simulations. For this experiment, we considered 100 users deployed over an area with 5 different locations, and energy loss rate $\beta = 0.2$. We consider $w_l = w_s = 0.5$ to emphasize equal weightage for both components. Rest of the parameters remain same as explained in Section \ref{sec:sim_params}.

In this setting, location attachments, and therefore, social links, are inferred by the algorithm automatically by monitoring patterns of co-location between nodes However, we can also preset the location attachment for the users based on their preferences. In the following, we present the comparison of the social context-aware algorithm (marked as $w/ social context$) over the mobility-aware algorithm (marked as $w/o social context$) with respect to the four metrics -- total network energy, total energy variation distance, number of P2P meetings, and number of nodes that reached energy balance.

\begin{figure*}[!ht]
        \centering
        \begin{subfigure}[b]{0.25\textwidth}
                \includegraphics[width=\textwidth]{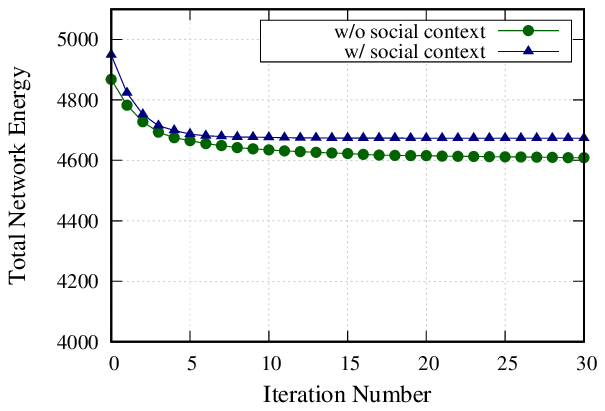}
                \caption{Total network energy.}
                \label{fig:1_sc_mob:tot_energy}
        \end{subfigure}%
        \begin{subfigure}[b]{0.25\textwidth}
                \includegraphics[width=\textwidth]{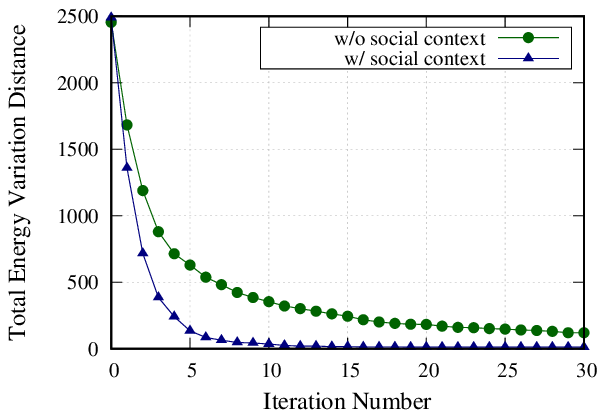}
                \caption{Total energy variation distance.}
                \label{fig:1_sc_mob:tot_var_dist}
        \end{subfigure}%
        \begin{subfigure}[b]{0.25\textwidth}
                \includegraphics[width=\textwidth]{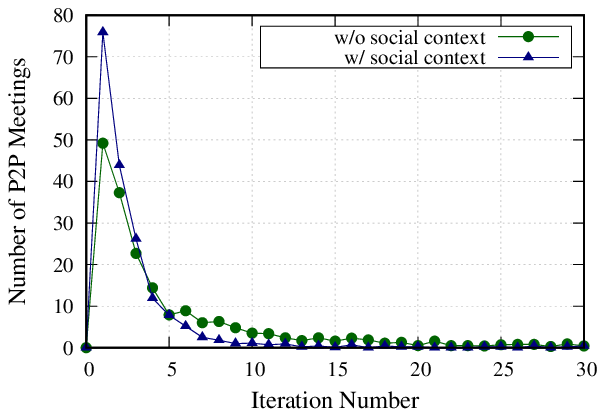}
                \caption{Number of P2P meetings.}
                \label{fig:1_sc_mob:num_meet}
        \end{subfigure}%
        \begin{subfigure}[b]{0.25\textwidth}
                \includegraphics[width=\textwidth]{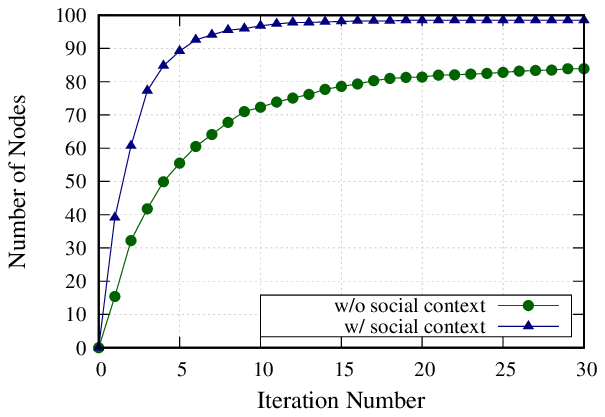}
                \caption{Nodes reached energy balance.}
                \label{fig:1_sc_mob:num_complete}
        \end{subfigure}%
        \caption{Performance results showing effects of social context in P2P-WPT $(\beta = 0.2, m = 100)$.}
\end{figure*}

In Figure \ref{fig:1_sc_mob:tot_energy}, the results for the total network energy is shown for $w/o social context$ and $w/ social context$. With the increase in iterations, the overall network energy reduces due to the energy loss during P2P-WPT between nodes. It can be seen from the results that in $w/ social context$, the overall network energy remains higher compared to $w/o social context$. This can be attributed to the fact that the energy loss in $w/ social context$ is less compared to $w/o social context$ method. Next, we compute the total energy variation distance for each iteration, and the results are shown in Figure \ref{fig:1_sc_mob:tot_var_dist}. We find that the total energy variation distance reduced when social context is considered. Also, in the initial iterations (1--10), the reduction in the variation distance in the $w/ social context$ method is much higher compared to the $w/o social context$ method. This can be attributed to the fact that $w/ social context$ method promotes higher number of P2P interactions during the initial iterations (1--5), which in turn helps in rapid reduction of the energy variation distance. Figure \ref{fig:1_sc_mob:num_meet} shows the results for the number of P2P interaction between the users, i.e., the number of selected pairs of peer, in each iteration. Consequently, the number of nodes which achieve the target energy balance is also higher in $w/ social context$ compared to $w/o social context$, as shown in Figure \ref{fig:1_sc_mob:num_complete}. As higher number of nodes start energy exchange in $w/ social context$, the total energy variation distance also decreases quickly. We also note that after each iteration, the total network energy remains higher in $P_{SC}$. Thus, we can infer that the overall energy balance in $w/ social context$ enhances compared to $w/o social context$. Additionally, the results shown in Figures \ref{fig:1_sc_mob:num_meet} and \ref{fig:1_sc_mob:num_complete} show that $w/ social context$ converges quickly compared to $w/o social context$.

\subsection{Effect of Social Relations in P2P-WPT}\label{sec:sec:eff_scr}
To motivate the effect of social relations in P2P-WPT, we evaluate and compare the performance of the social context and relation-aware peer selection method, named as $w/ social relations$ with the social context-aware method, named as $w/o social relations$. For this experiment also, we considered 100 users deployed over an area with 5 different locations, and energy loss rate $\beta = 0.2$. Here, we consider $w_l = w_s = w_e = 0.33$ to emphasize equal weightage for all three components. The movements of the users remain same as described in Section \ref{sec:sec:soc_cont}. Additionally, with the influence of the socially connected users, any user spends a random time in each location randomly distributed over $[0, f]$, where $f$ is the number of socially connected users at that location at that time. Rest of the parameters remain same as explained in Section \ref{sec:sim_params}.

\begin{figure*}[!ht]
        \centering
        \begin{subfigure}[b]{0.25\textwidth}
                \includegraphics[width=\textwidth]{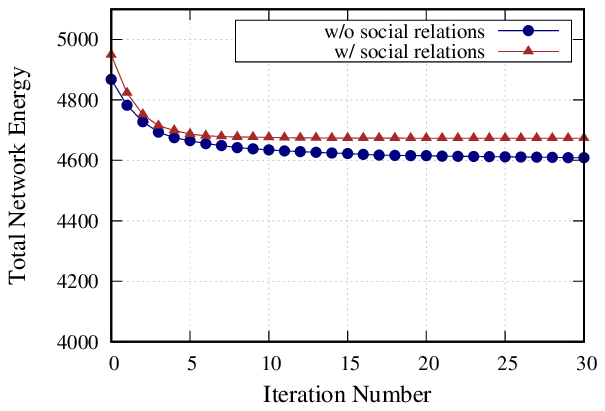}
                \caption{Total network energy.}
                \label{fig:2_sc_mob:tot_energy}
        \end{subfigure}%
        \begin{subfigure}[b]{0.25\textwidth}
                \includegraphics[width=\textwidth]{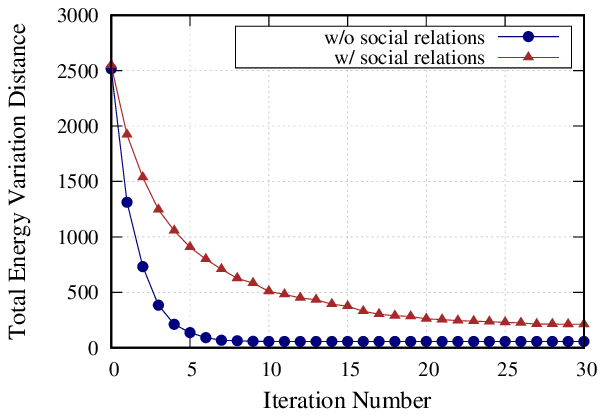}
                \caption{Total energy variation distance.}
                \label{fig:2_sc_mob:tot_var_dist}
        \end{subfigure}%
        \begin{subfigure}[b]{0.25\textwidth}
                \includegraphics[width=\textwidth]{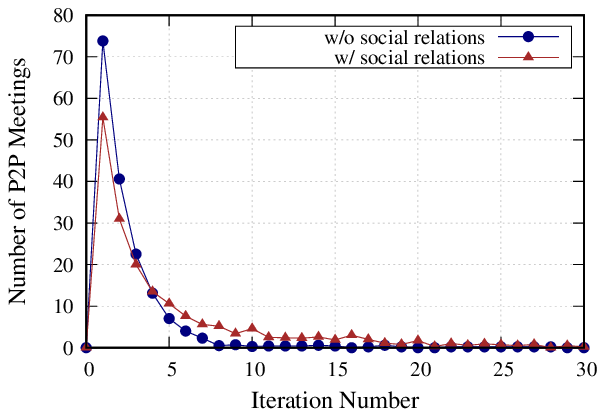}
                \caption{Number of P2P meetings.}
                \label{fig:2_sc_mob:num_meet}
        \end{subfigure}%
        \begin{subfigure}[b]{0.25\textwidth}
                \includegraphics[width=\textwidth]{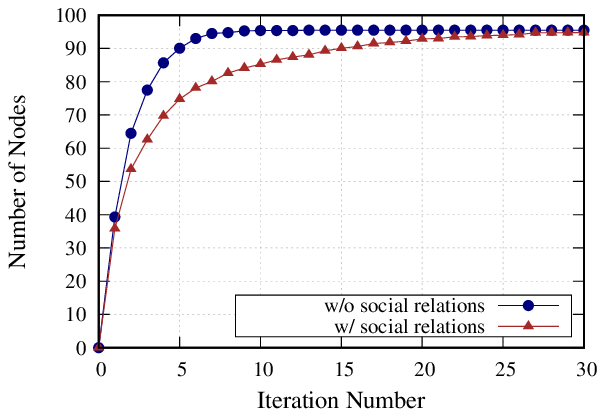}
                \caption{Nodes reached energy balance.}
                \label{fig:2_sc_mob:num_complete}
        \end{subfigure}%
        \caption{Performance results showing effects of social relations in P2P-WPT $(\beta = 0.2, m = 100)$.}
\end{figure*}

We discuss the results for $w/ social relations$ and $w/o social relations$ with respect to the four metrics -- total network energy, total energy variation distance, number of P2P meetings, and number of nodes that reached energy balance. Figure \ref{fig:2_sc_mob:tot_energy} and \ref{fig:2_sc_mob:tot_var_dist} present the total network energy and the energy variation distance, respectively. Whereas, Figure \ref{fig:2_sc_mob:num_meet} and \ref{fig:2_sc_mob:num_complete} shows the number of P2P interactions and the number of nodes achieving the target energy balance level. It can be seen from the results that in $w/ social relations$, the resulting network maintains higher total network energy compared to $w/o social relations$ method with an equal number of nodes reaching energy balance level. In case of the number of P2P interactions, $w/ social relations$ and $w/o social relations$ enable nearly equal number of iterations in iterations 2--5. In the next few iterations $w/ social relations$ enables higher number of P2P interactions between the users. As a result of this, the number of nodes which achieves energy balance increases gradually while keeping the energy loss minimum and the total remaining energy of the network higher. However, due to increased interactions during the initial iterations, the energy variation distance rapidly decreases in $w/o social relations$ and it remains less than that of $w/ social relations$. Therefore, we can conclude that $w/ social relations$ method can achieve better performance for total network energy with trade-off over the energy variation distance.

\section{Performance Comparison of MoSaBa with Alternative Approaches}\label{sec:Result_2}
In this section, we present the performance evaluation of the proposed MoSaBa method with the alternative approaches from the state-of-the-art. We first explain the chosen benchmarks, and then, discussed the results comparing with these approaches in detail.

\subsection{Benchmarks}
To compare the performance of the proposed mobility and social-aware energy balancing method $MoSaBa$, we consider three state-of-the-art methods as benchmarks -- $MobiWEB$ \cite{Ojha2021}, $P_{GO}$ \cite{Dhungana2019} and $P_{FT}$ \cite{Raptis2020}. The first benchmark $MobiWEB$ presents mobility-aware energy balancing for P2P-WPT as described by Ojha \textit{et al.} \cite{Ojha2021}. In $MobiWEB$, the algorithm has knowledge about the user mobility and leverages this information during peer selection for P2P-WPT. The peer selection process chooses the nodes with energy closest to the target energy balance level first, and then selects another node which has energy in the other side of target optimal energy level (compared to the first node). Based on this criteria nodes are chosen in pairs of peers and they participate in energy exchange until reaching the optimal energy level. $MobiWEB$ does not take into account any social information in its computation. The next benchmark $P_{GO}$ refers to greedy optimal energy balancing algorithm presented by Dhungana \textit{et al.} \cite{Dhungana2019}. In $P_{GO}$, pairs of peers are chosen such that they have energy closest to and in opposite side of the optimal average energy. Next, each such pair keeps exchanging energy until one of the nodes reaches the optimal energy level first. However, this method did not consider user mobility and rather selects nodes based on the energy condition only. The third benchmark is $P_{FT}$, which refers to the friend transfer protocol depicted by Raptis \textit{et al.} \cite{Raptis2020}. In this algorithm, two users exchange energy only in case they are social friends. In such interaction, the users splits their energy equally and exchanges energy until they reach that level. All three benchmarks consider the energy loss during P2P exchange. For all the benchmarks, we made few adjustments to enable fair comparison with the proposed method. For example, we consider each P2P-WPT exchange duration bounded by the inter-node meeting duration, which is decided randomly in the experiments. Also, in $P_{FT}$, the P2P interactions select nodes with energy value at the opposite side of the energy target level (not any two different energy values), as considered in the proposed method.

We choose these three benchmarks from the state-of-the-art, so as to cover the three different dimensions considered in our proposed methodology -- energy (current energy of any node), mobility and social aspects. $MobiWEB$ considers both mobility and energy aspects, whereas $P_{GO}$ and $P_{FT}$ considers the energy and social aspects respectively. Thereby, we believe these three methods will provide a fair benchmark for the evaluation of the proposed method, $MoSaBa$.

\begin{figure*}[ht]
        \centering
        \begin{subfigure}[b]{0.25\textwidth}
                \includegraphics[width=\textwidth]{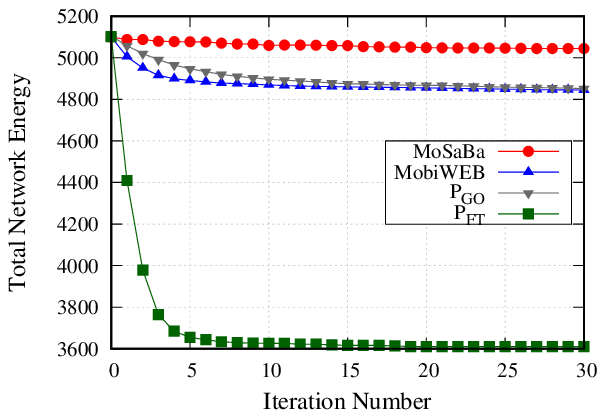}
                \caption{Total network energy.}
                \label{fig:3_all:tot_energy}
        \end{subfigure}%
        \begin{subfigure}[b]{0.25\textwidth}
                \includegraphics[width=\textwidth]{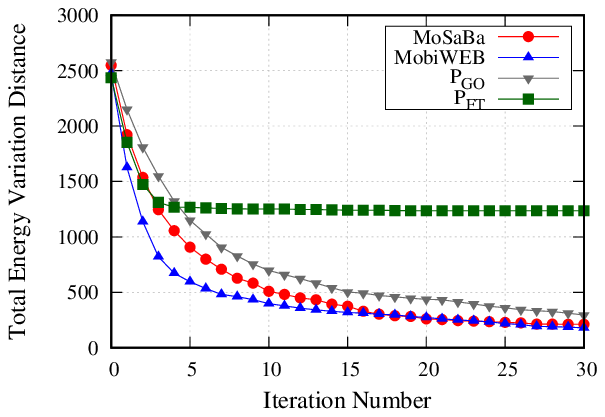}
                \caption{Total energy variation distance.}
                \label{fig:3_all:tot_var_dist}
        \end{subfigure}%
        \begin{subfigure}[b]{0.25\textwidth}
                \includegraphics[width=\textwidth]{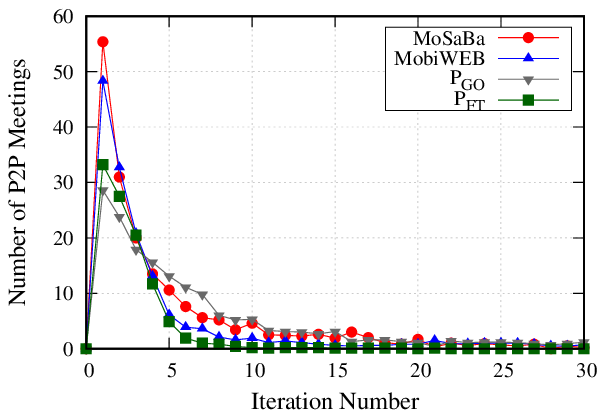}
                \caption{Number of P2P meetings.}
                \label{fig:3_all:num_meet}
        \end{subfigure}%
        \begin{subfigure}[b]{0.25\textwidth}
                \includegraphics[width=\textwidth]{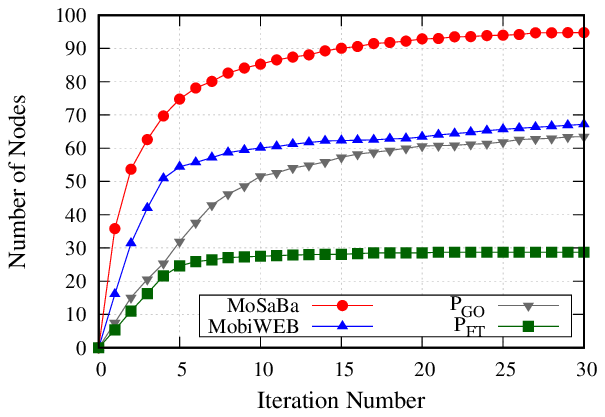}
                \caption{Nodes reached energy balance.}
                \label{fig:3_all:num_complete}
        \end{subfigure}%
        \caption{Performance comparison of $MoSaBa$ with the benchmark methods $(\beta = 0.2, m = 100)$.}
\end{figure*}

\subsection{Results}
In the following, we first present the results for all the methods with 100 nodes and energy loss rate $\beta$ = 0.2. Subsequently, in Section \ref{sec:sec:effect_nodes}, we discuss the results for all four metrics varying the number of nodes to 125 and 150. Then, in Section \ref{sec:sec:effect_beta}, we discuss the results for all four metrics with varying values of $\beta$ at 0.3 and 0.4. We also show the actual execution time for all the methods with respect to iterations.

\subsubsection{Total network energy}
In Figure \ref{fig:3_all:tot_energy}, we present the results for the total network energy at different iterations. Due to the effect of energy loss, with increase in iteration the total network energy reduces. From the results, we can find that the total network energy in the proposed $MoSaBa$ method remains higher compared to all the benchmarks. This attributes that the energy loss in the proposed method is lowest among all the methods. In $P_{FT}$, the energy loss is highest among all. This is due to the fact that in $P_{FT}$, the nodes exchanges energy with other nodes if they friend. As a result, the nodes may engage in exchanging higher amounts of energy, and subsequently, the energy loss also increases. On the other hand, in case other three methods, the energy exchange process is regulated by selecting the nodes with energy closest and opposite to the optimal energy level first. Due to such selection, the energy loss value remains lower. However, compared to $P_{GO}$, in the initial iterations (1--5), $MobiWEB$ has higher energy loss due to its higher number of P2P interactions in that duration (depicted in Figure \ref{fig:3_all:num_meet}). On the other hand, in $MoSaBa$, although the number of P2P interactions in the initial iterations are higher, the mobility and social-aware peer selection ensures low energy loss, and thus, the total network energy remains higher than others.

\subsubsection{Total energy variation distance} 
Figure \ref{fig:3_all:tot_var_dist} depicts the results for the total energy variation distance at different iteration. From the results it is evident that during the initial iterations (1--5), the energy variation distance in all the methods decreases rapidly according to the number of interactions in each method. Both $MoSaBa$ and $MobiWEB$ enable higher number of P2P interactions in the initial iterations. However, the energy variation distance in $MoSaBa$ remains higher compared to $MobiWEB$ only during the initial iterations, and finally, by the middle of the experiment (\~iteration 15) the both the methods reach nearly equal value of energy variation distance. For the other two methods the energy variation distance remains higher. Among these two methods, in $P_{FT}$, the energy variation distance value remains nearly same for most of experiments (especially after iteration 10) due to nearly no interactions between nodes in those iterations. Therefore, from these results, we can understand that the proposed method is able to achieve a good quality of energy balance in terms of minimizing both energy variation distance as well as the energy loss.

\subsubsection{Number of P2P meetings}
In Figure \ref{fig:3_all:num_meet}, the number of P2P interactions between the nodes in different iterations are noted. As discussed previously, the number of P2P interactions in the initial iterations (1--5) are very high in all the methods. However, after that such events reduces -- as some of the nodes reach energy balance, the effective number of nodes which can be considered as a potential peer reduces. We note from the results that the proposed method $MoSaBa$ enables P2P interaction opportunity for the users in rest of the experiment especially during iterations 5--20, $MoSaBa$ provides higher number of P2P interactions compared to that of the benchmarks. The number of nodes which achieves the target energy balance level also increases rapidly in this duration. Therefore, we can infer that our proposed method converges quickly as well as provides better P2P-WPT opportunities.

\subsubsection{Number of nodes that reach energy balance} 
In Figure \ref{fig:3_all:num_complete}, we provide the number of nodes which achieves the target energy balance level. From the results, we can find that only in $MobiWEB$ and in the proposed method $MoSaBa$, more that 50\% nodes reach energy balance within iteration 4. However, in $MoSaBa$, energy balance of more than 70\% nodes is achieved within iteration 5. Also, the proposed method can reach highest number of nodes reaching the energy balance level while maintaining a lower energy loss and energy variation distance. From these facts, we can conclude that the proposed method, $MoSaBa$, has better quality of energy balance among all the methods.

\begin{figure*}[!ht]
        \centering
        \begin{subfigure}[b]{0.25\textwidth}
                \includegraphics[width=\textwidth]{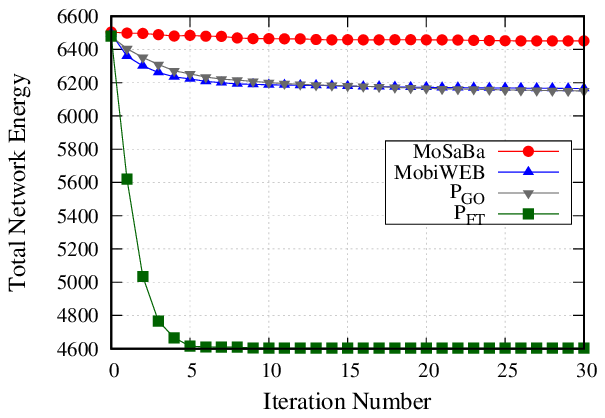}
                \caption{Total network energy.}
                \label{fig:5_nodes_125:tot_energy}
        \end{subfigure}%
        \begin{subfigure}[b]{0.25\textwidth}
                \includegraphics[width=\textwidth]{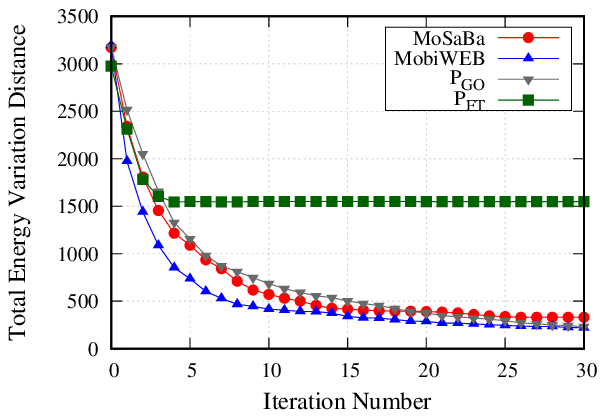}
                \caption{Total energy variation distance.}
                \label{fig:5_nodes_125:tot_var_dist}
        \end{subfigure}%
        \begin{subfigure}[b]{0.25\textwidth}
                \includegraphics[width=\textwidth]{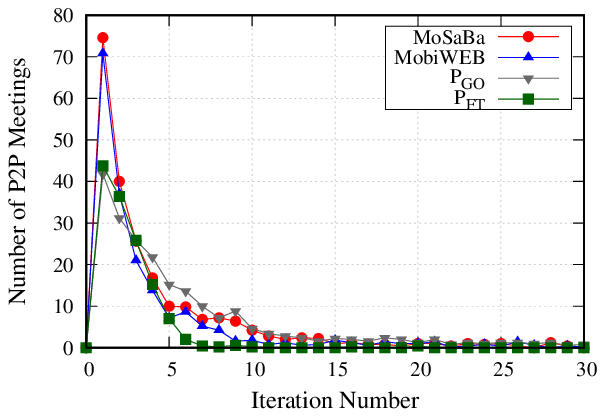}
                \caption{Number of P2P meetings.}
                \label{fig:5_nodes_125:num_meet}
        \end{subfigure}%
        \begin{subfigure}[b]{0.25\textwidth}
                \includegraphics[width=\textwidth]{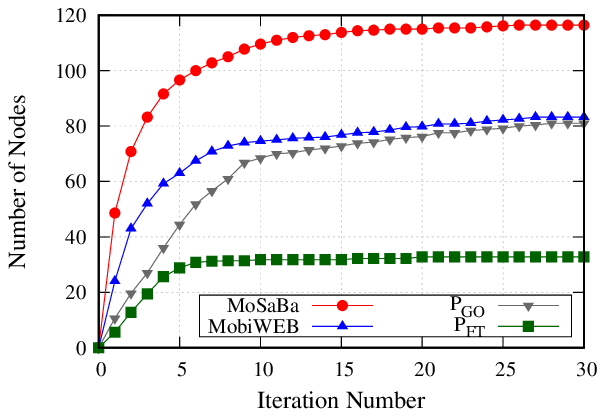}
                \caption{Nodes reached energy balance.}
                \label{fig:5_nodes_125:num_complete}
        \end{subfigure}%
        \caption{Performance comparison of $MoSaBa$ with the benchmark methods $(\beta = 0.2, m = 125)$.}
\end{figure*}

\begin{figure*}[!ht]
        \centering
        \begin{subfigure}[b]{0.25\textwidth}
                \includegraphics[width=\textwidth]{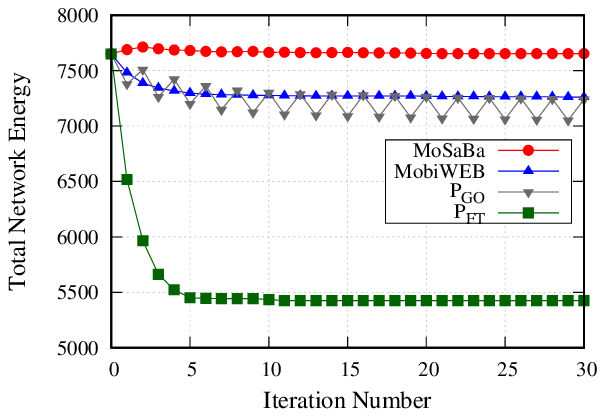}
                \caption{Total network energy.}
                \label{fig:5_nodes_150:tot_energy}
        \end{subfigure}%
        \begin{subfigure}[b]{0.25\textwidth}
                \includegraphics[width=\textwidth]{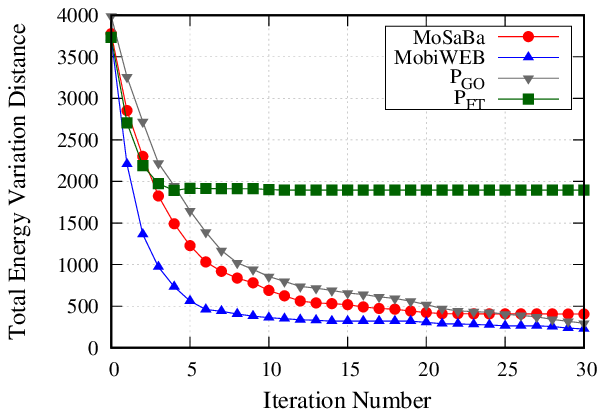}
                \caption{Total energy variation distance.}
                \label{fig:5_nodes_150:tot_var_dist}
        \end{subfigure}%
        \begin{subfigure}[b]{0.25\textwidth}
                \includegraphics[width=\textwidth]{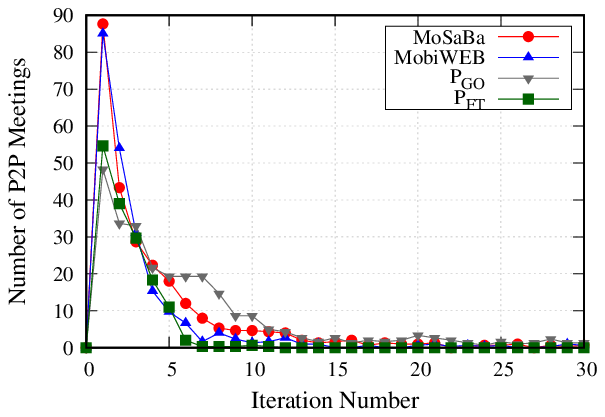}
                \caption{Number of P2P meetings.}
                \label{fig:5_nodes_150:num_meet}
        \end{subfigure}%
        \begin{subfigure}[b]{0.25\textwidth}
                \includegraphics[width=\textwidth]{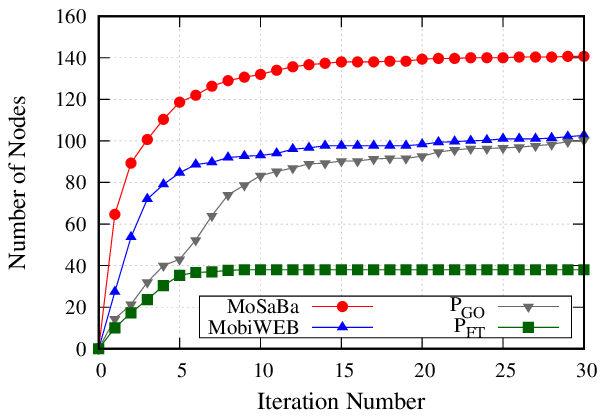}
                \caption{Nodes reached energy balance.}
                \label{fig:5_nodes_150:num_complete}
        \end{subfigure}%
        \caption{Performance comparison of $MoSaBa$ with the benchmark methods $(\beta = 0.2, m = 150)$.}
\end{figure*}

\subsubsection{Effect of number of nodes}\label{sec:sec:effect_nodes}
Here, we present the results for all the evaluation metrics varying the values of number of nodes and discuss their effects on these metrics. In Figure \ref{fig:5_nodes_125:tot_energy} and \ref{fig:5_nodes_150:tot_energy}, we present the results for total network energy for 125 and 150 nodes, respectively. These results show similar patterns of behavior among the methods as depicted for 100 nodes. However, due to presence of increased number of nodes, the overall energy in the network is higher. From the results, we can find that the proposed method $MoSaBa$ maintains the lowest values of energy loss and thus, the total network energy in this case remains highest. With increase in the number of nodes, the methods are able to find increased number of P2P opportunities. As a result, in our proposed method, the energy loss rate remains lowest among all the methods. 

Figures \ref{fig:5_nodes_125:tot_var_dist} and \ref{fig:5_nodes_150:tot_var_dist} present the results for total energy variation distance achieved in the different methods for 125 and 150 nodes, respectively. The results for this metric also follows similar pattern as with 100 nodes. However, with the increase in number of nodes, the total energy variation distance value increases for all the nodes. In the proposed method, the comparative increase in the energy variation distance between 100-125 to 125-150 is lower compared to rest of the methods. Although, it is important to note that the total network energy remains highest in the proposed method. Thus, we can argue that proposed method $MoSaBa$ is better in conserving the network energy.

In Figure \ref{fig:5_nodes_125:num_meet} and \ref{fig:5_nodes_150:num_meet}, we present the number of P2P meetings for 125 and 150 nodes, respectively. It is important to note from the results that the percentage of nodes engaged in P2P meetings during the initial iterations (1--5) has increased more in the proposed method compared to the benchmarks. This is due to the fact that with an increase in number of nodes, the proposed method is able to find more pairs of peers in the initial iterations. Therefore, we can infer that with an increase in the number of nodes, the proposed method converges quickly compared to the benchmarks.

Figures \ref{fig:5_nodes_125:num_complete} and \ref{fig:5_nodes_150:num_complete} show the results for the number of nodes which achieve energy balancing for 125 and 150 nodes respectively. Here, the percentage of nodes which reached energy balancing increases slightly for all the methods. Similar to other metrics, in this case also, the increase in potential number of peers has resulted in increased number of nodes that achieves energy balance level. Therefore, considering all other results, we can infer that the proposed P2P-WPT method, $MoSaBa$, is scalable with number of nodes.

\begin{figure*}[!ht]
        \centering
        \begin{subfigure}[b]{0.25\textwidth}
                \includegraphics[width=\textwidth]{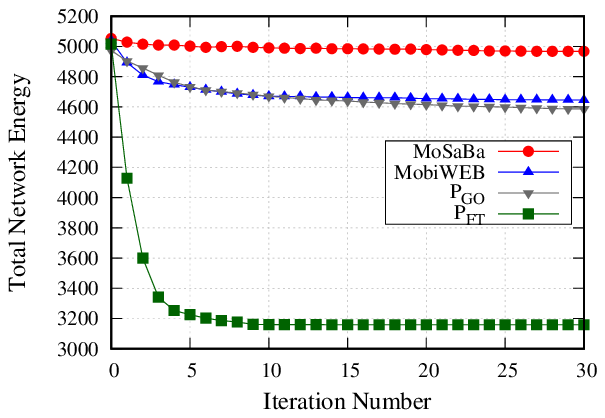}
                \caption{Total network energy.}
                \label{fig:4_beta_0.3:tot_energy}
        \end{subfigure}%
        \begin{subfigure}[b]{0.25\textwidth}
                \includegraphics[width=\textwidth]{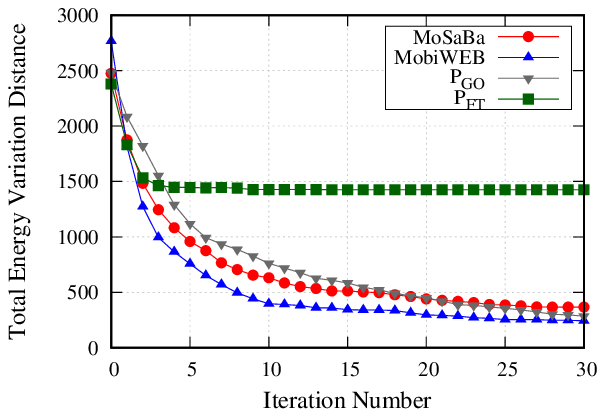}
                \caption{Total energy variation distance.}
                \label{fig:4_beta_0.3:tot_var_dist}
        \end{subfigure}%
        \begin{subfigure}[b]{0.25\textwidth}
                \includegraphics[width=\textwidth]{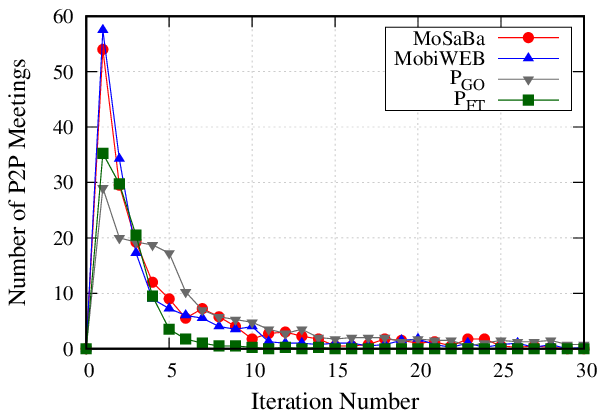}
                \caption{Number of P2P meetings.}
                \label{fig:4_beta_0.3:num_meet}
        \end{subfigure}%
        \begin{subfigure}[b]{0.25\textwidth}
                \includegraphics[width=\textwidth]{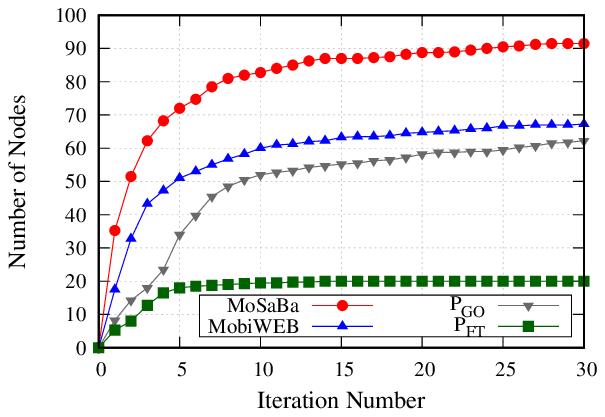}
                \caption{Nodes reached energy balance.}
                \label{fig:4_beta_0.3:num_complete}
        \end{subfigure}%
        \caption{Performance comparison of $MoSaBa$ with the benchmark methods $(\beta = 0.3, m = 100)$.}
\end{figure*}

\begin{figure*}[!ht]
        \centering
        \begin{subfigure}[b]{0.25\textwidth}
                \includegraphics[width=\textwidth]{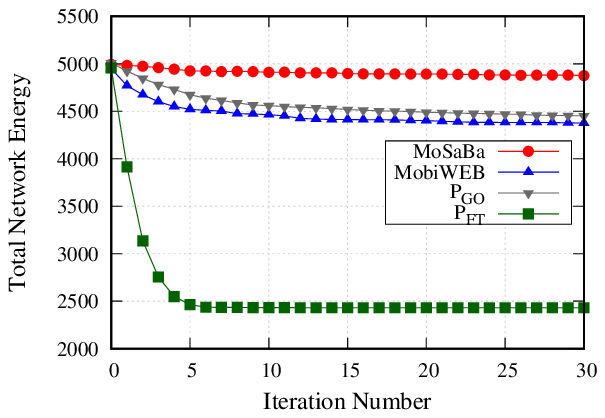}
                \caption{Total network energy.}
                \label{fig:4_beta_0.4:tot_energy}
        \end{subfigure}%
        \begin{subfigure}[b]{0.25\textwidth}
                \includegraphics[width=\textwidth]{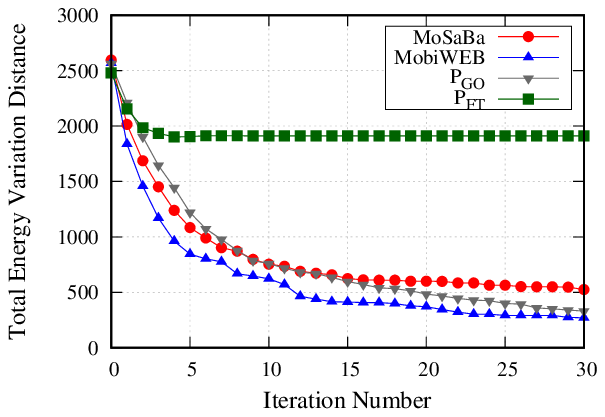}
                \caption{Total energy variation distance.}
                \label{fig:4_beta_0.4:tot_var_dist}
        \end{subfigure}%
        \begin{subfigure}[b]{0.25\textwidth}
                \includegraphics[width=\textwidth]{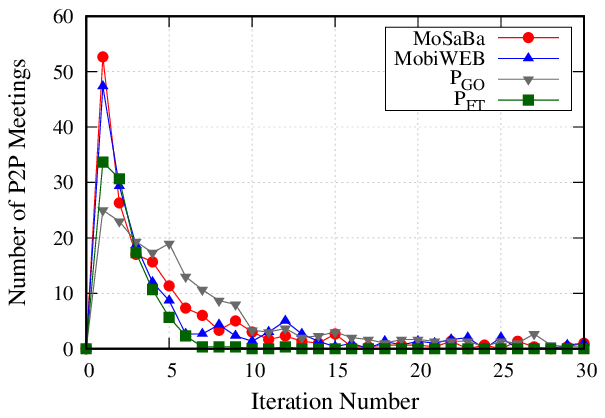}
                \caption{Number of P2P meetings.}
                \label{fig:4_beta_0.4:num_meet}
        \end{subfigure}%
        \begin{subfigure}[b]{0.25\textwidth}
                \includegraphics[width=\textwidth]{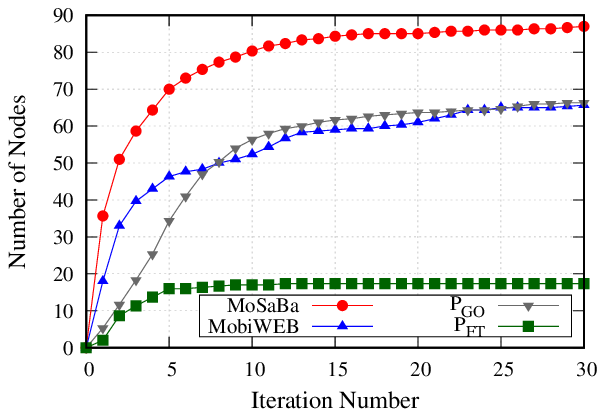}
                \caption{Nodes reached energy balance.}
                \label{fig:4_beta_0.4:num_complete}
        \end{subfigure}%
        \caption{Performance comparison of $MoSaBa$ with the benchmark methods $(\beta = 0.4, m = 100)$.}
\end{figure*}

\subsubsection{Effect of energy loss rate}\label{sec:sec:effect_beta}
In Figures \ref{fig:4_beta_0.3:tot_energy} and \ref{fig:4_beta_0.4:tot_energy}, we present the results for the total network energy metric for $\beta$ value of 0.3 and 0.4, respectively. With increase in the energy loss rate, all of the methods experience higher energy loss. However, the energy loss remains lower in the proposed method compared to the benchmarks, where the nodes lose energy quickly. Also, the rate of decrease of the total network energy is highest in case of $P_{FT}$, as in this case, the nodes engage in higher amount of energy exchange compared to other methods, and subsequently, the energy loss increase rapidly in this method. Both $MobiWEB$ and $P_{GO}$, which facilitates energy based peer selection, results in comparatively lower amount of energy loss than $P_{FT}$. However, in case of the proposed method, $MoSaBa$, the energy loss remains lowest due to incorporating both energy and location/social-aware decision making. 

Figure \ref{fig:4_beta_0.3:tot_var_dist} and \ref{fig:4_beta_0.4:tot_var_dist} shows the results for the total energy variation distance for $\beta$ value of 0.3 and 0.4, respectively. With increase in the energy loss rate, the difference in energy variation distance between the methods reduces. We can also notice that the initial iterations (1--5) show comparatively less decrease in the variation distance. This is due to the fact that higher rate of energy loss has increased the individual energy variation distance or the nodes reach the desired value of energy balance with increased number of iteration. Therefore, as expected, higher loss affects the energy balancing process by having higher energy variation between nodes and thereby delaying the energy balance process.

In Figure \ref{fig:4_beta_0.3:num_meet} and \ref{fig:4_beta_0.4:num_meet}, we present the results for the number of P2P meetings in case of $\beta$ value of 0.3 and 0.4, respectively. The number of P2P interactions enabled by the different methods significantly in the initial iterations (1--5) and also reduces slightly during the early to middle iterations (5--15). Thus, the results for this metric also suggests the slower progress of energy balancing process.

Figure \ref{fig:4_beta_0.3:num_complete} and \ref{fig:4_beta_0.4:num_complete} shows the results for number of nodes that reach the energy balancing level for $\beta$ value of 0.3 and 0.4, respectively. The increase of energy loss rate affects the number of nodes that achieve energy balancing for all the methods. However, as the proposed method $MoSaBa$ is able to find comparatively higher number of P2P opportunities (and also for more prolonged period), the number of nodes achieving energy balance remains significantly higher compared to the benchmarks. However, with increase of energy loss rate, the energy variation distance also increases (although the difference becomes less in initial iterations rather than in later iterations). Therefore, we can infer that the proposed method can provide comparatively better quality of energy balancing for increased energy loss rate, with performance trade-off in energy variation distance.

\subsubsection{Execution Time}\label{sec:sec:exec_time}
\begin{figure}[ht]
        \centering
        \begin{subfigure}[b]{0.5\linewidth}
                \includegraphics[width=\textwidth]{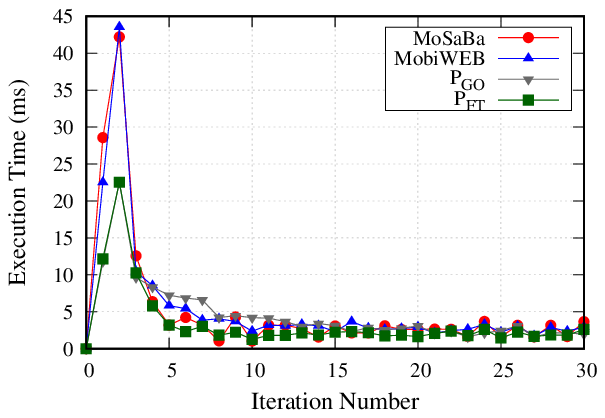}
                \caption{$m = 100$.}
                \label{fig:3_all:exec_time100}
        \end{subfigure}%
        \begin{subfigure}[b]{0.5\linewidth}
                \includegraphics[width=\textwidth]{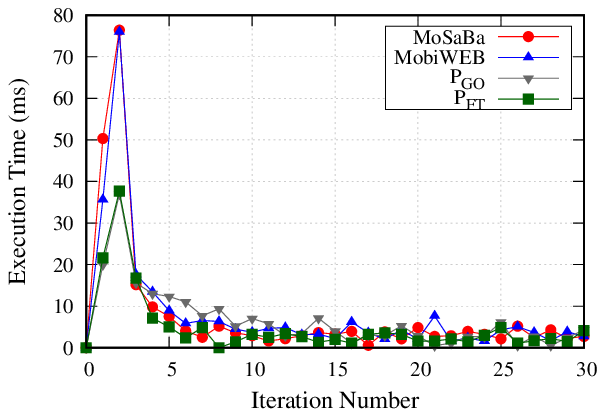}
                \caption{$m = 150$.}
                \label{fig:3_all:exec_time150}
        \end{subfigure}%
        \caption{Comparison of execution time for all the methods $(\beta = 0.2)$.}
\end{figure}

In Figure \ref{fig:3_all:exec_time100} and \ref{fig:3_all:exec_time150}, we show the actual CPU execution time in each iteration of the simulation for the proposed and benchmark methods. From the results, it is evident that the execution time remains higher in the early iterations (1--5) for all the methods. This is due to the fact that most of the P2P meetings happen in those iterations leading to higher value of execution time. Consequently, the execution time reduces in the later iterations (6--30) with the reduction of P2P meetings. Also, both $MoSaBa$ and $MobiWEB$ results in higher execution time compared to $P_{GO}$ and $P_{FT}$. Such behavior is attributed to the higher number of P2P meetings executed in both $MoSaBa$ and $MobiWEB$ compared to the other two methods. Similarly, with the increase of the number of nodes from 100 to 150, the execution times for all the methods increase with higher number of P2P meetings. Therefore, we can infer that the execution time is correlated with the number of P2P meetings in each iteration.

From the results of execution time, we see that the average decision delay in the busiest iterations, i.e., in iterations 1--5 when most of the P2P meetings are performed, is $< 1$ $ms$ only. For a large set of nodes, there might be cases where multiple nodes are with same energy level -- i.e., more than one node with energy closest to as well as belongs in the \textit{same side} of $\overline{E}^*$. In such cases, we need to prioritize any one among these nodes (for simplicity, we can choose first occurring node), and the other nodes (having same energy level) will experience a decision delay of $< 1$ $ms$ only. Therefore, in the current scenario and given conditions, we can apply the proposed heuristic for scenarios with high number of nodes.

\section{Conclusion}\label{sec:Conclu}
In this paper, we present a wireless crowd charging method, named \textit{MoSaBa}, by exploiting the mobility information and social relations of the users. In contrast to the existing literature, we also consider more fine-grained and realistic assumption in the underlying P2P-WPT process. In \textit{MoSaBa}, the mobility prediction for the users is performed by using the Markov predictor. Thereafter, we quantify and compute the impact of social contexts as well as social relations on the peer selection process. Based on these information, we explore different pairs of peers for energy exchange such that the crowd achieves energy balance faster while maintaining an overall low energy loss. Simulation-based results show that the energy balance quality improves in the proposed method -- due to low energy loss, the overall energy of the crowd after energy balancing remains higher. Compared to the existing works, in \textit{MoSaBa}, more number of users successfully achieve the desired energy balance level. Also, the proposed method has faster convergence compared to the benchmarks. Therefore, we can conclude that the proposed method is able to achieve performance trade-offs between energy-efficiency, energy balance quality and convergence time. In future, we plan to extend the proposed method in a real-world application with heterogeneous devices. We also plan to incorporate more additional energy loss for the users (due to different activities e.g. mobility, communication) and maintain more fine-grained energy expenses.

\section*{Acknowledgment}
This work was carried out during the tenure of an ERCIM `Alain Bensoussan’ Fellowship Programme of the first author.

\bibliographystyle{IEEEtran}
\bibliography{Mobility_p2pWPT_ext}


\end{document}